\newif\ifacm
\acmtrue

\newif\ifcomment
\commenttrue

\newif\ifcameraready
\camerareadyfalse

\newif\ifwatermark
\watermarkfalse

\ifwatermark
      \usepackage{draftwatermark}
      \SetWatermarkScale{.7}
      \SetWatermarkText{\shortstack{In Submission:\\Do Not Distribute}}
      \SetWatermarkColor[rgb]{1,0.88,0.88}
\fi

\ifcomment
    \newcounter{MVNumberOfComments}
    \newcounter{HCNumberOfComments}
    \newcounter{SMNumberOfComments}
    \newcounter{HFNumberOfComments}
    \stepcounter{MVNumberOfComments}
    \stepcounter{HCNumberOfComments}
    \stepcounter{SMNumberOfComments}
    \stepcounter{HFNumberOfComments}
    \newcommand{\mvnote}[1]{\textcolor{blue}{\small \bf [MV\#\arabic{MVNumberOfComments}\stepcounter{MVNumberOfComments}: #1]}}
    \newcommand{\hcnote}[1]{\textcolor{red}{\small \bf [HC\#\arabic{HCNumberOfComments}\stepcounter{HCNumberOfComments}: #1]}}
    \newcommand{\smnote}[1]{\textcolor{green}{\small \bf [SM\#\arabic{SMNumberOfComments}\stepcounter{SMNumberOfComments}: #1]}}
    \newcommand{\hfnote}[1]{\textcolor{orange}{\small \bf [HF\#\arabic{HFNumberOfComments}\stepcounter{HFNumberOfComments}: #1]}}

    \newcommand{\NOTE}[1]
    {
      {\footnotesize\it
        \begin{center}
          \begin{tabular}{|c|}
           \hline
            \parbox{0.85\columnwidth}{
              \medskip
              #1
              \medskip} \\
            \hline
          \end{tabular}
        \end{center}
        }
    }
\else
    \newcommand\mvnote[1]{}
    \newcommand\hcnote[1]{}
    \newcommand\smnote[1]{}
    \newcommand\hfnote[1]{}
    \newcommand\NOTE[1]{}
\fi

\newcommand{\eg}{{e.g.,}\xspace}
\newcommand{\ie}{{\it i.e.,}\xspace}

\newcounter{NumTakeaways}
\stepcounter{NumTakeaways}

\ifacm
    \documentclass[sigconf]{acmart}

\else 
    \documentclass[10pt, conference, letterpaper]{IEEEtran}
\fi

\ifacm
\copyrightyear{2021}
\acmYear{2021}
\setcopyright{acmlicensed}\acmConference[IMC '21]{ACM Internet Measurement Conference}{November 2--4, 2021}{Virtual Event, USA}
\acmBooktitle{ACM Internet Measurement Conference (IMC '21), November 2--4, 2021, Virtual Event, USA}
\acmPrice{15.00}
\acmDOI{10.1145/3487552.3487847}
\acmISBN{978-1-4503-9129-0/21/11}

\usepackage{gensymb}
\usepackage{booktabs}  
\usepackage{pbox}
\usepackage{subcaption}
\usepackage{epsfig,endnotes}
\usepackage{epstopdf}
\usepackage{times}
\usepackage{color}
\usepackage{colortbl} 
\usepackage{graphicx}
\usepackage{xspace}
\usepackage{enumitem}
\usepackage{multirow}
\usepackage{tikz}
\usepackage[ruled,vlined,linesnumbered]{algorithm2e}
\usepackage{lmodern}
\usepackage[T1]{fontenc}
\usepackage[utf8]{inputenc}
\usepackage{array}
\usepackage[normalem]{ulem}
\usepackage{siunitx}
\usepackage{cancel}
\hypersetup{draft}
\usepackage{soul}
\soulregister\cite7
\soulregister\ref7
\soulregister\pageref7

\else
\usepackage{cite}
\usepackage{amsmath,amssymb,amsfonts}
\usepackage{algorithmic}
\usepackage{graphicx}
\usepackage{textcomp}
\usepackage{xcolor}
\def\BibTeX{{\rm B\kern-.05em{\sc i\kern-.025em b}\kern-.08em
    T\kern-.1667em\lower.7ex\hbox{E}\kern-.125emX}}

\usepackage{gensymb}
\usepackage{booktabs}  
\usepackage{pbox}
\usepackage{subcaption}
\usepackage{epsfig,endnotes}
\usepackage{epstopdf}
\usepackage{times}
\usepackage{color}
\usepackage{colortbl} 
\usepackage{graphicx}
\usepackage{xspace}
\usepackage{enumitem}
\usepackage{multirow}
\usepackage{tikz}
\usepackage[ruled,vlined,linesnumbered]{algorithm2e}
\usepackage{lmodern}
\usepackage[T1]{fontenc}
\usepackage[utf8]{inputenc}
\usepackage{array}
\usepackage[normalem]{ulem}
\usepackage{siunitx}
\usepackage{cancel}
\hypersetup{draft}

\fi

\newcolumntype{L}[1]{>{\raggedright\let\newline\\\arraybackslash\hspace{0pt}}m{#1}}
\newcolumntype{C}[1]{>{\centering\let\newline\\\arraybackslash\hspace{0pt}}m{#1}}
\newcolumntype{R}[1]{>{\raggedleft\let\newline\\\arraybackslash\hspace{0pt}}m{#1}}

\begin{document}

\title{Can You See Me Now? A Measurement Study of Zoom, Webex, and Meet}


\ifacm
\settopmatter{authorsperrow=4}
    \author{Hyunseok Chang}
    \affiliation{%
      \institution{Nokia~Bell Labs}
      \city{Murray Hill}
      \state{NJ}
      \country{USA}
    }
    \author{Matteo Varvello}
    \affiliation{%
      \institution{Nokia~Bell Labs}
      \city{Murray Hill}
      \state{NJ}
      \country{USA}
    }
    \author{Fang Hao}
    \affiliation{%
      \institution{Nokia~Bell Labs}
      \city{Murray Hill}
      \state{NJ}
      \country{USA}
    }
    \author{Sarit Mukherjee}
    \affiliation{%
      \institution{Nokia~Bell Labs}
      \city{Murray Hill}
      \state{NJ}
      \country{USA}
    }
\else
    \author{Paper \#XXX}
\fi 

%

\ifacm
    \begin{abstract}
Since the outbreak of the COVID-19 pandemic, videoconferencing has become the default mode of communication in our daily lives at homes, workplaces and schools, and it is likely to remain an important part of our lives in the post-pandemic world.  Despite its significance, there has not been any systematic study characterizing the user-perceived performance of existing videoconferencing systems other than anecdotal reports.  In this paper, we present a detailed measurement study that compares three major videoconferencing systems: Zoom, Webex and Google Meet.  Our study is based on 48 hours' worth of more than 700 videoconferencing sessions, which were created with a mix of emulated videoconferencing clients deployed in the cloud, as well as real mobile devices running from a residential network.  We find that the existing videoconferencing systems vary in terms of geographic scope, which in turns determines streaming lag experienced by users. We also observe that streaming rate can change under different conditions (e.g., number of users in a session, mobile device status, etc), which affects user-perceived streaming quality.  Beyond these findings, our measurement methodology can enable reproducible benchmark analysis for any types of comparative or longitudinal study on available videoconferencing systems.
\end{abstract}

\ccsdesc[500]{Networks~Network measurement}
\ccsdesc[300]{Information systems~Collaborative and social computing systems and tools}
\ccsdesc[300]{Networks~Cloud computing}

    \maketitle
\else 
    \maketitle
    \begin{abstract}
Since the outbreak of the COVID-19 pandemic, videoconferencing has become the default mode of communication in our daily lives at homes, workplaces and schools, and it is likely to remain an important part of our lives in the post-pandemic world.  Despite its significance, there has not been any systematic study characterizing the user-perceived performance of existing videoconferencing systems other than anecdotal reports.  In this paper, we present a detailed measurement study that compares three major videoconferencing systems: Zoom, Webex and Google Meet.  Our study is based on 48 hours' worth of more than 700 videoconferencing sessions, which were created with a mix of emulated videoconferencing clients deployed in the cloud, as well as real mobile devices running from a residential network.  We find that the existing videoconferencing systems vary in terms of geographic scope, which in turns determines streaming lag experienced by users. We also observe that streaming rate can change under different conditions (e.g., number of users in a session, mobile device status, etc), which affects user-perceived streaming quality.  Beyond these findings, our measurement methodology can enable reproducible benchmark analysis for any types of comparative or longitudinal study on available videoconferencing systems.
\end{abstract}

\fi 

\section{Introduction}
\label{sec:intro}

There is no doubt that the outbreak of the COVID-19 pandemic has fundamentally changed our daily lives.  Especially with everyone expected to practice physical distancing to stop the spread of the pandemic, various online communication tools have substituted virtually all sorts of in-person interactions.   As the closest form of live face-to-face communication in a pre-pandemic world, videoconferencing has practically become the default mode of communication (\eg an order-of-magnitude increase in videoconferencing traffic at the height of the pandemic
~\cite{nokiablog,webexblog,feldmann2020}).
 Thanks to its effectiveness and reliability, video communication is likely to remain an important part of our lives even in the post-pandemic world~\cite{ozimek20,cdc20}.

Despite the critical role played by existing videoconferencing systems in our day-to-day communication, there has not been any systematic study on quantifying their performance and Quality of Experience (QoE).  There is no shortage of anecdotal reports and discussions in terms of the usability, video quality, security, and client resource usage of individual systems. To the best of our knowledge, however, no scientific paper has yet investigated the topic thoroughly with a sound measurement methodology that is applicable across multiple available systems.  Our main contribution in this paper addresses this shortcoming.

In this paper, we shed some light on the existing videoconferencing ecosystems by characterizing their infrastructures as well as their performance from a user's QoE perspective. To this end, we have devised a measurement methodology which allows us to perform controlled and reproducible benchmarking of  videoconferencing systems by leveraging a mix of emulated videoconferencing clients deployed in the cloud, as well as real mobile devices running from a residential network.  We provide the detailed description of our methodology as well as the open-source tools we used (Sections~\ref{sec:arch} and~\ref{sec:res}), so that anyone can replicate our testbed to repeat or further extend our benchmark scenarios.  Driven by our methodology, we investigate three popular videoconferencing systems on the market: Zoom, Webex and Google Meet (Meet for short).  Each of these platforms provides a free-tier plan as well as paid subscriptions, but we focus on their \emph{free-tier plans} in our evaluation. Given these three systems, we conduct measurement experiments which take a combined total of 48 videoconferencing hours over more than 700 sessions, with 200 VM hours rented from 12 geographic locations and 18 hours of two Android phones hooked up at one location.  Our findings include: 

\vspace{0.1in}
\noindent\textbf{Finding-1.} In the US, typical streaming lag experienced by users is \SIrange[range-phrase=--,range-units=single]{20}{50}{\ms} for Zoom, \SIrange[range-phrase=--,range-units=single]{10}{70}{\ms} for Webex, and \SIrange[range-phrase=--,range-units=single]{40}{70}{\ms} for Meet.  This lag largely reflects the geographic separation of users (\eg US-east vs.~US-west). In case of Webex, all sessions created in the US appear to be relayed via its infrastructure in US-east.  This causes the sessions among users in US-west to be subject to artificial detour, inflating their streaming lag.

\vspace{0.1in}
\noindent\textbf{Finding-2.} Zoom and Webex are characterized by a US-based infrastructure. It follows that sessions created in Europe experience higher lag than those created in the US (\SIrange[range-phrase=--,range-units=single]{90}{150}{\ms} for Zoom, and \SIrange[range-phrase=--,range-units=single]{75}{90}{\ms} for Webex).  On the other hand, the sessions created in Europe on Meet exhibit smaller lag (\SIrange[range-phrase=--,range-units=single]{30}{40}{\ms}) due to its cross-continental presence including Europe.

\vspace{0.1in}
\noindent\textbf{Finding-3.} All three systems appear to optimize their streaming for low-motion videos (e.g., a single-person view with a stationary background). Thus high-motion video feeds (e.g., dynamic scenes in outdoor environments) experience non-negligible QoE degradation compared to typical low-motion video streaming.

\vspace{0.1in}
\noindent\textbf{Finding-4.} Given the same camera resolution, Webex sessions exhibit the highest traffic rate for multi-user sessions.  Meet exhibits the most dynamic rate changes across different sessions, while Webex maintains virtually constant rate across sessions.

\vspace{0.1in}
\noindent\textbf{Finding-5.} Videoconferencing is an expensive task for mobile devices, requiring at least 2--3 full cores to work properly. Meet is the most bandwidth-hungry client, consuming up to one GB per hour, compared to Zoom's gallery view that only requires \SI{175}{MB} per hour. We estimate that one hour's videoconferencing can drain up to 40\% of a low-end phone's battery, which can be reduced to about 20--30\% by turning off the onboard camera/screen and relying only on audio. All videoconferencing clients scale well with the number of call participants, thanks to their UI which only displays a maximum of four users at a time.

\vspace{0.1in}
The rest of the paper is organized as follows. We start by introducing related works in Section~\ref{sec:related}.  We then present our measurement methodology in Section~\ref{sec:arch}, and describe the measurement experiments and our findings in detail in Sections~\ref{sec:res}--\ref{sec:res:usage}.  We conclude in Section~\ref{sec:disc} by discussing several research issues.

\section{Related Work}
\label{sec:related}
Despite the prevalence of commercial videoconferencing systems~\cite{singh2020updated}, no previous work has directly compared them with respect to their infrastructures and end-user QoE, which is the main objective of this paper. The recent works by~\cite{sandervideo} and \cite{macmillan21} investigate the network utilization and bandwidth sharing behavior of existing commercial videoconferencing systems based on controlled network conditions and client settings.
Several works propose generic solutions to improve videoconferencing. For example, Dejavu~\cite{hu2019dejavu} offers up to 30\% bandwidth reduction, with no impact on QoE, by leveraging the fact that recurring videoconferencing sessions have lots of similar content, which can be cached and re-used across sessions. Salsify~\cite{salsify} relies on tight integration between a video codec and a network transport protocol to dynamically adjust video encodings to changing network conditions.

\begin{table}[t]
\small
\centering
\begin{tabular}{|c|c|c|}\hline
\textbf{Videoconferencing system}  & \textbf{Low quality}       & \textbf{High quality}\\\hline
Zoom~\cite{zoom_req}    & \multicolumn{2}{c|}{\SI{600}{Kbps}}   \\\hline
Webex~\cite{webex_req}  & \SI{500}{Kbps}    & \SI{2.5}{Mbps}    \\\hline
Meet~\cite{meet_req}    & \SI{1}{Mbps}      & \SI{2.6}{Mbps}    \\\hline
\end{tabular}
\caption{Minimum bandwidth requirements for one-on-one calls.}
\label{tab:bw}
\vspace{-8ex}
\end{table}

As a consequence of the COVID-19 pandemic, the research community has paid more attention to the impact of videoconferencing systems on the quality of education~\cite{correia2020evaluating, zou2020covid, parra2021evaluating}. As educational studies, these works rely on usability analysis and student surveys. In contrast, our work characterizes QoE performance of the videoconferencing systems using purely objective metrics.

Videoconferencing operators do not provide much information about their system, \eg footprint and encoding strategies. One common information reported by each operator is the minimum bandwidth requirements for one-on-one calls (Table~\ref{tab:bw}).  The results from our study are not only consistent with these requirements, but also cover more general scenarios such as multi-party sessions.




 
\section{Benchmarking Design}
\label{sec:arch}
In this section, we describe the benchmarking tool we have designed to study  existing commercial videoconferencing systems. We highlight key design goals for the tool first, followed by associated challenges, and then describe how we tackle the challenges in our design.

A videoconferencing system is meant to be used by end-users in mobile or desktop environments that are equipped with a camera and a microphone.  When we set out to design our benchmarking tool for such systems, we identify the following design goals.

\vspace{0.05in}
\noindent\textbf{(D1) Platform compliance:} We want to run our benchmark tests using \emph{unmodified} official videoconferencing clients with \emph{full audiovisual capabilities}, so that we do not introduce any artifact in our evaluation that would be caused by client-side incompatibility or deficiency.

\vspace{0.05in}
\noindent\textbf{(D2) Geo-distributed deployment:} To evaluate web-scale videoconferencing services in realistic settings, we want to collect data from \emph{geographically-distributed} clients.

\vspace{0.05in}
\noindent\textbf{(D3) Reproducibility:} We want to leverage \emph{a controlled, reproducible client-side environment}, so that we can compare available videoconferencing systems side-by-side.

\vspace{0.05in}
\noindent\textbf{(D4) Unified evaluation metrics:} We want to evaluate different videoconferencing platforms based on \emph{unified metrics} that are applicable across all the platforms.
\vspace{0.1in}

It turns out that designing a benchmarking tool that meets all the stated goals is challenging because some of these goals are in fact conflicting. For example, while geographically-dispersed public clouds can provide distributed vantage point environments (\textbf{D2}), cloud deployments will not be equipped with necessary sensory hardware (\textbf{\xcancel{D1}}).  Crowd-sourced end-users can feed necessary audiovisual data into the videoconferencing systems (\textbf{D1}), but benchmarking the systems based on unpredictable human behavior and noisy sensory data will not give us objective and reproducible comparison results (\textbf{\xcancel{D3}}).  On the other hand, some goals such as \textbf{(D3)} and \textbf{(D4)} go hand in hand.  Unified evaluation metrics alone are not sufficient for comparative analysis if reproducibility of client-side environments is not guaranteed.  At the same time, reproducibility would not help much without platform-agnostic evaluation metrics.

\subsection{Design Approach}
Faced with the aforementioned design goals and challenges, we come up with a videoconferencing benchmark tool that is driven by three main ideas: (i) client emulation, (ii) coordinated client deployments, and (iii) platform-agnostic data collection.

\vspace{0.1in}
\noindent
\textbf{Client emulation.} One way to circumvent the requirement for sensory devices in videoconferencing clients is to \emph{emulate} them. Device emulation also means that the sensory input to a videoconferencing client would be completely under our control, which is essential to reproducible and automated benchmarking.  To this end, we leverage \emph{loopback pseudo devices} for audio/video input/output.  In Linux, \texttt{snd-aloop} and \texttt{v4l2loopback} modules allow one to set up a virtual soundcard device and a virtual video device, respectively. Once activated these loopback devices appear to videoconferencing clients as standard audio/video devices, except that audiovisual data is not coming from a real microphone or a capture card, but is instead sourced by other applications. In our setup we use \texttt{aplay} and \texttt{ffmpeg} to replay audio/video files into these virtual devices.  The in-kernel device emulation is completely transparent to the clients, thus no client-side modification is required.

Another aspect of client emulation is client UI navigation. Each videoconferencing client has client-specific UI elements for interacting with a videoconferencing service, such as logging in, joining/leaving a meeting, switching layouts, etc. We automate UI navigation of deployed clients by emulating various input events (e.g., keyboard typing, mouse activity, screen touch) with OS-specific tools (e.g., \texttt{xdotool} for Linux, and \texttt{adb-shell} for Android). For each videoconferencing system, we script the entire workflow of its client.

\vspace{0.1in}
\noindent\textbf{Coordinated client deployments.} Fully-emulated clients allow us to deploy the clients in public clouds and/or mobile testbeds for automated testing. The fact that we control audiovisual data feed for the clients as well as their UI navigation provides unique opportunities for us to gain, otherwise difficult to obtain, insights into the videoconferencing systems under test.  For example, one client can be injected with a video feed with specific patterns (e.g., periodic ON/OFF signals, or high-/low-motion videos), and other clients receive the feed through a videoconferencing service. By comparing the injected feed and received feeds, we can evaluate different videoconferencing services. We can easily coordinate the activities of multiple participants in a given conferencing session to facilitate our analysis (e.g., only one user's screen is active at a time).

\vspace{0.1in}
\noindent
\textbf{Platform-agnostic data collection.}  Even with client emulation and coordination, the closed nature of the existing videoconferencing systems (e.g., proprietary client software and end-to-end encryption) poses as a hurdle to comparing the systems with objective and unified metrics.  That has led us to perform data collection in a \emph{platform-agnostic} fashion as follows.

First, we derive some of the evaluation metrics from network-level monitoring and measurements. For example, we measure streaming lag by correlating packet timestamps on sender-side and on receiver-side. That way, we can evaluate the videoconferencing infrastructures without being influenced by specifics in client deployments.  This, however, requires accurate clock synchronization among deployed clients.  Fortunately, major public clouds already provide dedicated time sync services for tenant workloads with their own stratum-1 clock sources~\cite{awstimesync,azuretimesync}.

In order to supplement network-based metrics with user-perceived quality metrics, we record videoconferencing sessions from individual participants' perspective, and assess the quality of recorded audios/videos across different platforms. While Zoom provides a local recording option for each participant, other services like Webex or Meet only allow a meeting host to record a session.  In the end, we adopt a desktop recording approach as a platform-agnostic measure.  We run a videoconferencing client in full screen mode, and use \texttt{simplescreenrecorder} to record the desktop screen with audio, within a cloud VM itself.

Finally, we also evaluate the videoconferencing systems from their clients' resource-utilization perspectives, which is particularly important for mobile devices.  While these metrics can be influenced by client implementation, we believe that platform-driven factors (e.g., audio/video codecs) may play a bigger role.

\begin{figure}[t]
\center
\includegraphics[width = 0.9\linewidth,trim = 0mm 5mm 0mm 0mm, clip=true]{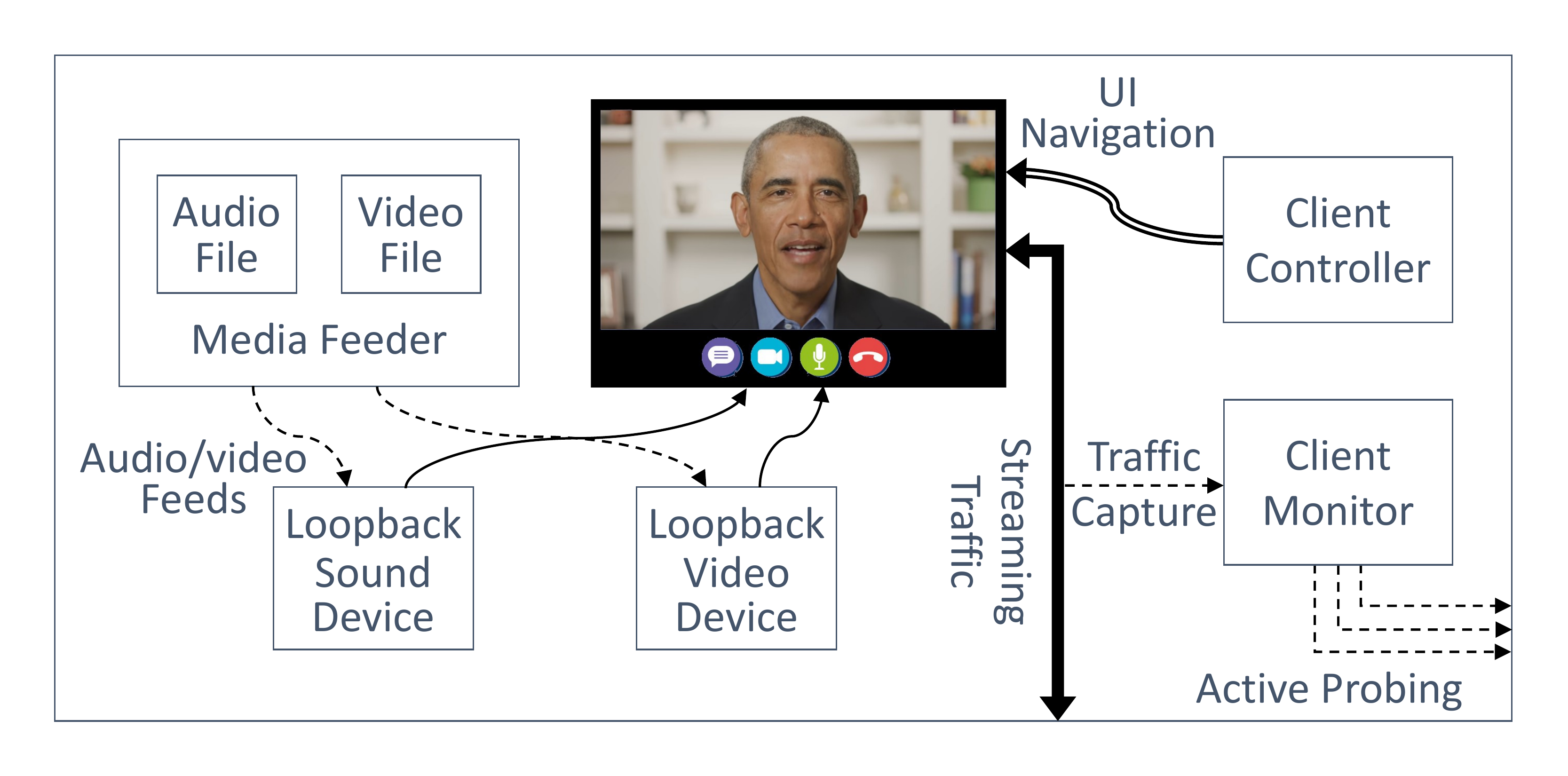}
\caption{Cloud VM in a fully-emulated environment.}
\label{fig:emul}
\vspace{-3ex}
\end{figure}

\subsection{Deployment Targets}
\label{sec:arch:pipeline}
Based on the design approach described above, we deploy emulated videoconferencing clients on a group of cloud VMs and Android mobile phones.  Each of the cloud VMs hosts a videoconferencing client in a fully emulated setting to generate (via emulated devices) and/or receive a streaming feed, while Android devices only receive feeds from videoconferencing systems without device emulation.\footnote{While Android devices can generate sensory data from their onboard camera/microphone, we mostly do not use them for reproducible benchmarking.} In the following, we describe each of these deployment targets in more details.

\vspace{0.1in}
\noindent\textbf{Cloud VM.} A cloud VM runs a videoconferencing client on a remote desktop in a fully-emulated environment.  It consists of several components as shown in Fig.~\ref{fig:emul}. \emph{Media feeder} replays audio and video files into corresponding loopback devices.  The audio and video files are either synthetically created, or extracted individually from a video clip with sound.  \emph{Client monitor} captures incoming/outgoing videoconferencing traffic with \texttt{tcpdump}, and dumps the trace to a file for offline analysis, as well as processes it on-the-fly in a separate ``active probing'' pipeline. In this pipeline, it discovers streaming service endpoints (IP address, TCP/UDP port) from packet streams, and performs round-trip-time (RTT) measurements against them.  We use \texttt{tcpping} for RTT measurements because ICMP pings are blocked at the existing videoconferencing infrastructures. \emph{Client controller} replays a platform-specific script for operating/navigating a client, including launch, login, meeting-join/-leave and layout change.

In order to host the cloud VMs, a public cloud must meet the following requirements. First, the cloud must \emph{not} be used to operate the videoconferencing systems under test. For example, the majority of Zoom infrastructure is known to be hosted at AWS cloud~\cite{zoomaws}. If we run our emulated clients in the same AWS environment, Zoom will be heavily favored in our evaluation due to potential intra-cloud network optimization. To prevent such bias, we exclude any public cloud being used by the videoconferencing systems we tested.  The cloud must also have reasonably wide geographic coverage. In the end we choose Azure cloud~\cite{azure} as our benchmarking platform.

\vspace{0.1in}
\noindent\textbf{Android devices.} We use Samsung Galaxy S10 and J3 phones, representative of both \textit{high-end} and \textit{low-end} devices (Table~\ref{tab:testbed}).  The battery of the J3 is connected to a Monsoon power meter~\cite{monsoon} which produces fine-grained battery readings.  Both devices are connected to a Raspberry Pi (via WiFi to avoid USB noise on the power readings) which is used to automate Android UI navigation and to monitor their resource utilization (\eg CPU usage). Both tasks are realized via Android Debugging Bridge (\texttt{adb}).  The phones connect to the Internet over a fast WiFi -- with a symmetric upload and download bandwidth of \SI{50}{Mbps}. Each device connects to its own WiFi realized by the Raspberry Pi, so that traffic can be easily isolated and captured for each device.

\begin{table}[t]
\footnotesize
\centering
\begin{tabular}{ccccccc}
\hline
    {\bf Name} &  {\bf Android Ver.}   & {\bf CPU Info} & {\bf Memory}  & {\bf Screen Resolution} \\
\hline
    Galaxy J3 & 8  & Quad-core  & 2GB  & 720x1280  \\
    Galaxy S10 & 11 & Octa-core & 8GB  & 1440x3040 \\
\hline
\end{tabular}
\caption{Android devices characteristics.} 
\vspace{-0.1in}
\label{tab:testbed}
\vspace{-0.1in}
\end{table}

\section{Quality of User Experience}
\label{sec:res}
In this section, we present QoE analysis results from our benchmark analysis of three major videoconferencing systems: Zoom, Webex and Meet.
The experiments were conducted from 4/2021 to 5/2021.

\subsection{Cloud VM Setup}
Each cloud VM we deploy has 8 vCPUs (Intel Xeon Platinum 8272CL with 2.60GHz), 16GB memory and 30GB SSD. We make sure that the provisioned VM resources are sufficient for all our benchmark tests, which involve device emulation, videoconferencing, traffic monitoring and desktop recording. The screen resolution of the VM's remote desktop is set to 1900$\times$1200.  We use the native Linux client for Zoom (v5.4.9 (57862.0110)), and the web client for Webex and Meet since they do not provide a native Linux client.

\subsection{Streaming Lag}
\label{sec:res:qoe:lag}

\begin{figure}[t]
\centering
\includegraphics[width = 0.8\linewidth,trim = 0mm 1mm 0mm 10mm, clip=true]{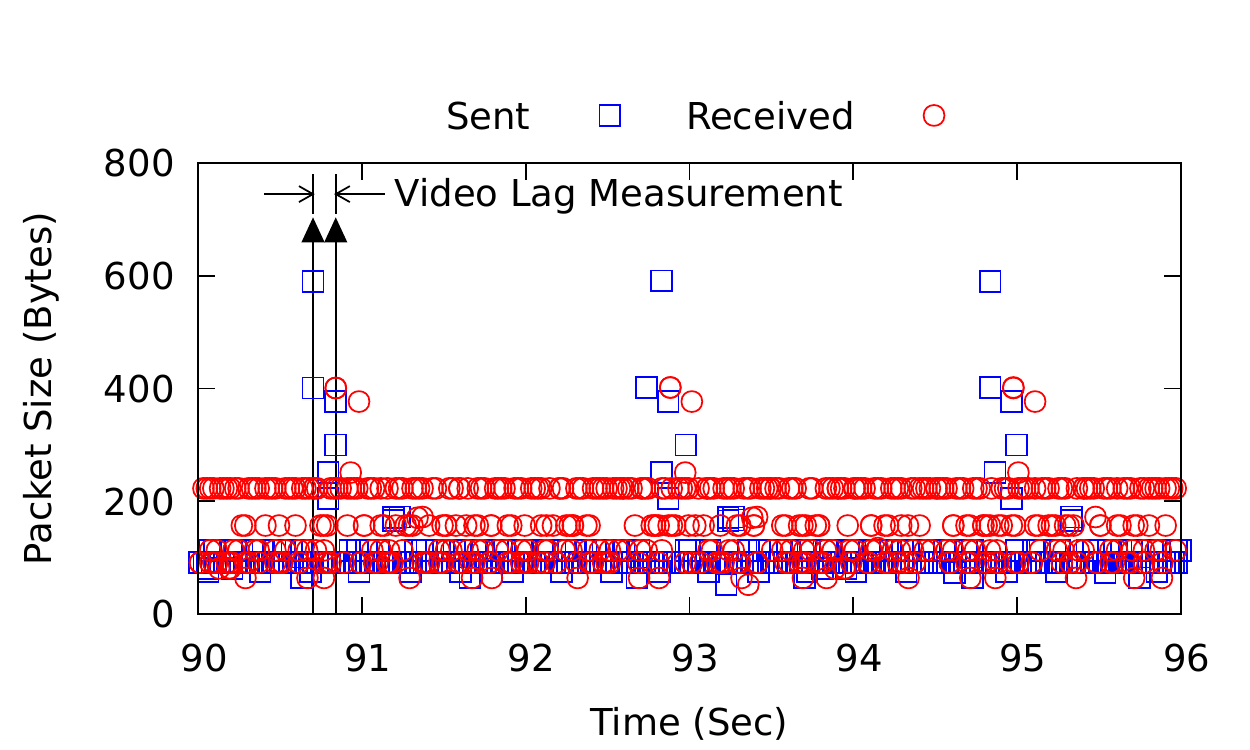}
\caption{Video lag measurement.}
\label{fig:pkt_viz}
\vspace{-2ex}
\end{figure}

First we evaluate streaming lag experienced by users (i.e., time delay between audio/video signals ingested by one user and those received by another). While measuring streaming lag in real-life videoconferencing is difficult, our emulated clients with synchronized clocks allow us to quantify the lags precisely. We purposefully set the video screen of a meeting host to be a blank-screen with periodic flashes of an image (with two-second periodicity), and let other users join the session with no audio/video of their own. Using such a bursty, one-way video feed allows us to easily determine the timing of sent/received video signals from network traffic monitoring.

For example, Fig.~\ref{fig:pkt_viz} visualizes the packet streams observed on the meeting host (sender) and another user (receiver). The squares represent the sizes of packets sent by a meeting host, and the circles show the sizes of packets received by a user. As expected there are periodic spikes of ``big'' packets ($>$200 bytes) that match periodic video signals sent and received.  The first big packet that appears after more than a second-long quiescent period indicates the arrival of a non-blank video signal.  We measure streaming lag between the meeting host and the other participant with the time shift between the first big packet on sender-side and receiver-side. Admittedly, this network-based metric discounts any potential delay caused by a receiver-side client (e.g., due to stream buffering/decoding). However, it is effective to evaluate and compare lags induced by streaming infrastructures and their geographic coverage.

\begin{table}[t]
\footnotesize
\centering
\begin{tabular}{|C{1.1cm}|C{2.0cm}|C{1.8cm}|C{0.9cm}|} \hline
Region & Location & Name & Count \\\hline\hline
\multirow{5}{*}{US} & Iowa & US-Central & 1\\
 & Illinois & US-NCentral & 1 \\
 & Texas & US-SCentral & 1 \\
 & Virginia & US-East & 2 \\
 & California & US-West & 2 \\\hline
\multirow{7}{*}{Europe} & Switzerland & CH & 1\\
 & Denmark & DE & 1\\
 & Ireland & IE & 1\\
 & Netherlands & NL & 1\\
 & France & FR & 1\\
 & London, UK & UK-South & 1\\
 & Cardiff, UK & UK-West & 1\\\hline
\end{tabular}
\caption{VM locations/counts for streaming lag testing.}
\label{tab:azureloc}
\vspace{-5ex}
\end{table}

\begin{figure}[t]
\centering
\includegraphics[width = 0.8\linewidth,trim = 0mm 9mm 0mm 0mm, clip=true]{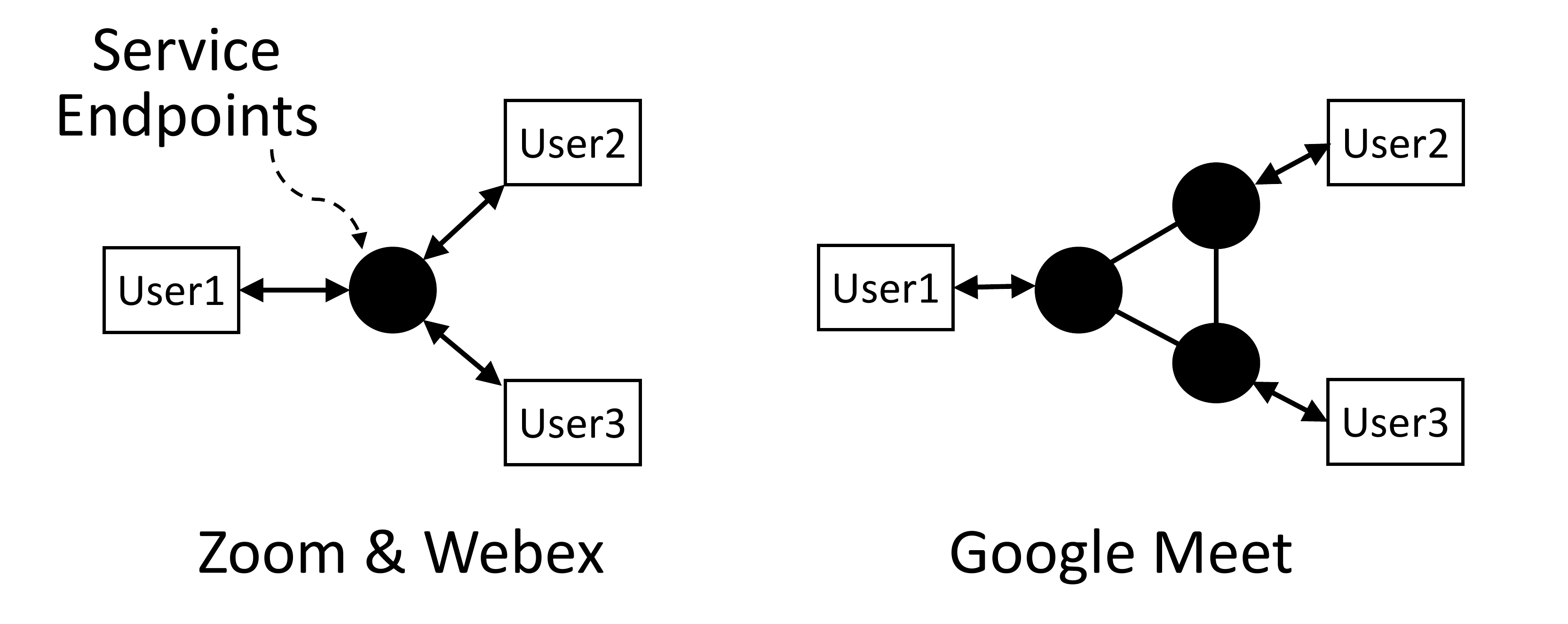}
\caption{Videoconferencing service endpoints.}
\label{fig:topo}
\end{figure}

In the first set of experiments we deploy seven VMs in the US, as indicated by the ``VM count'' field in Table~\ref{tab:azureloc}.  We create a meeting session with one VM (in either US-east or US-west) designated as a meeting host, which then broadcasts periodic video signals to the other six participating VMs for two minutes before terminating the session. We collect 35-40 lag measurements from each participant during the session. For more representative sampling, we create 20 such meeting sessions with the same meeting host. Thus in the end we have a total of 700-800 lag measurements from each of the six VMs for a particular meeting host.  We repeat this experiment on Zoom, Webex and Meet.  In the second set of experiments we use seven VMs deployed in Europe, as shown in Table~\ref{tab:azureloc}, and redo the above experiments with meeting hosts in UK-west and Switzerland.


\begin{figure*}[ht]
\centering
\begin{subfigure}{0.32\linewidth}
\centering
\includegraphics[width = 1.0\textwidth,trim = 0mm 0mm 0mm 15mm, clip=true]{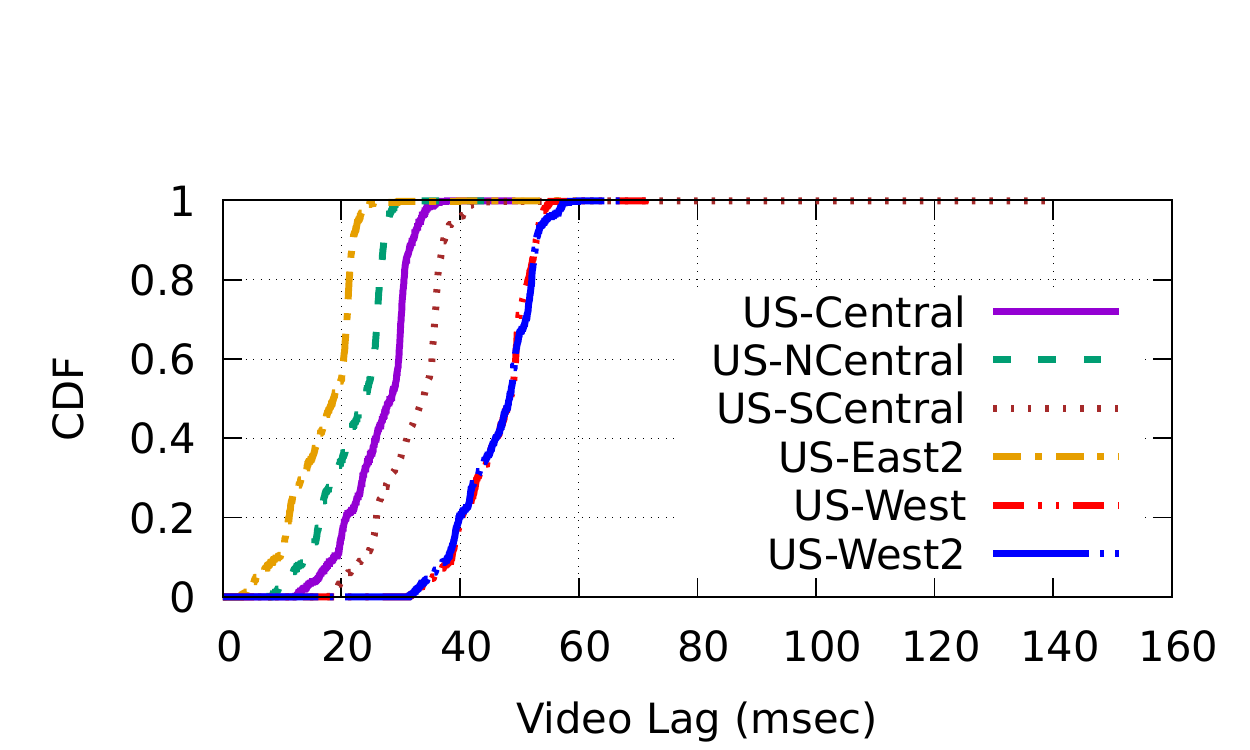}
\caption{Zoom}
\label{fig:lag_useast:zoom}
\end{subfigure}
\begin{subfigure}{0.32\linewidth}
\centering
\includegraphics[width = 1.0\textwidth,trim = 0mm 0mm 0mm 15mm, clip=true]{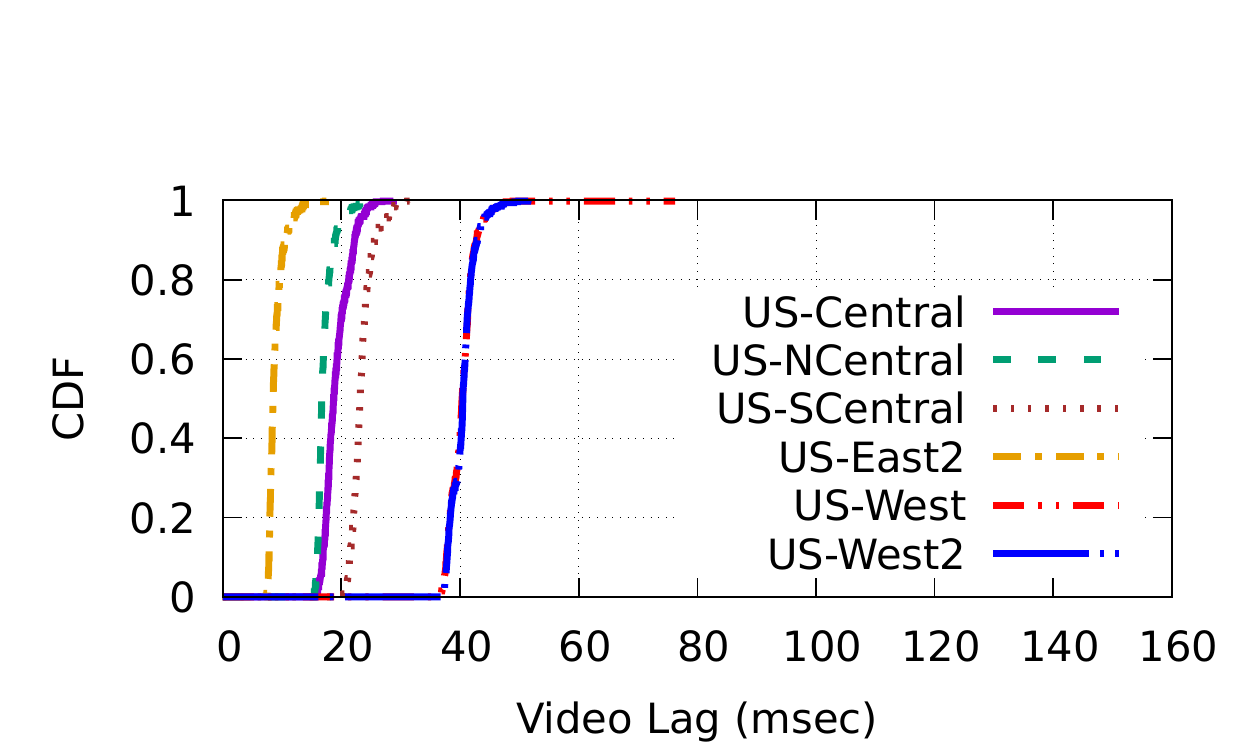}
\caption{Webex}
\label{fig:lag_useast:webex}
\end{subfigure}
\begin{subfigure}{0.32\linewidth}
\centering
\includegraphics[width = 1.0\textwidth,trim = 0mm 0mm 0mm 15mm, clip=true]{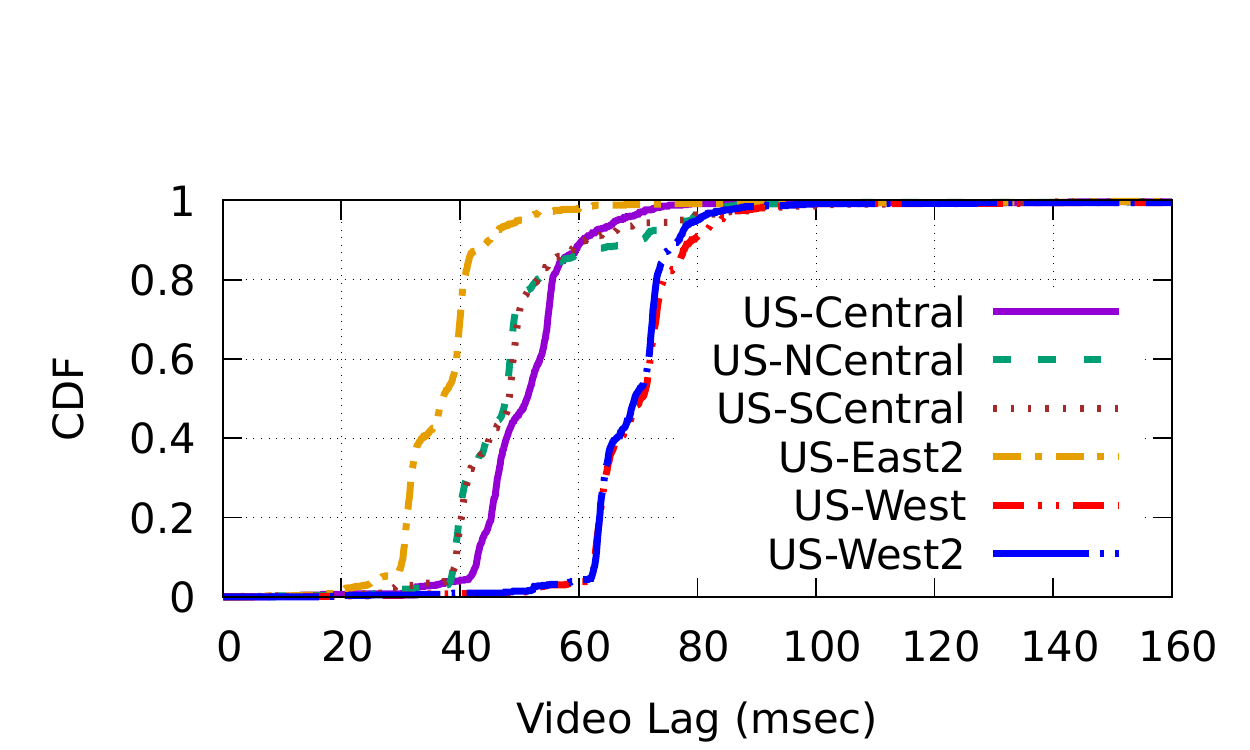}
\caption{Meet}
\label{fig:lag_useast:goog}
\end{subfigure}
\vspace{-1ex}
\caption{CDF of streaming lag: meeting host in US-east.}
\label{fig:lag_useast}
\vspace*{-2ex}
\end{figure*}

\begin{figure*}[ht]
\centering
\begin{subfigure}{0.32\linewidth}
\centering
\includegraphics[width = 1.0\textwidth,trim = 0mm 0mm 0mm 15mm, clip=true]{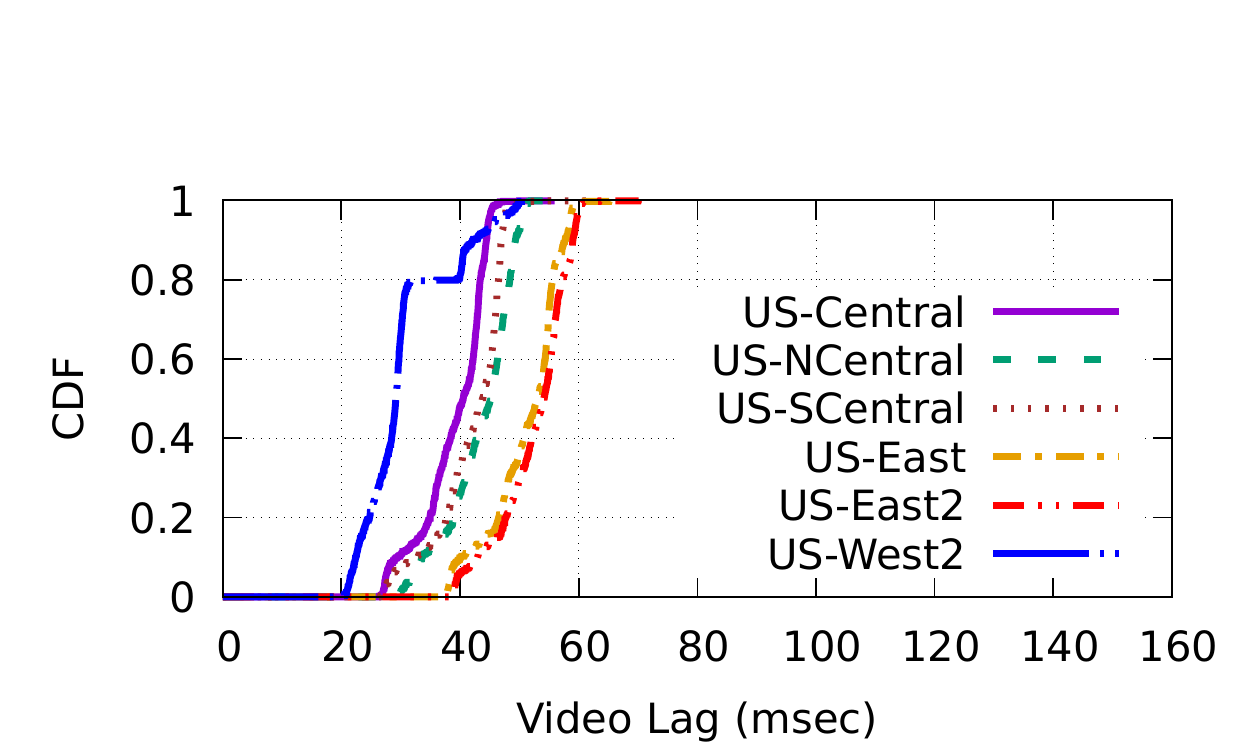}
\caption{Zoom}
\label{fig:lag_uswest:zoom}
\end{subfigure}
\begin{subfigure}{0.32\linewidth}
\centering
\includegraphics[width = 1.0\textwidth,trim = 0mm 0mm 0mm 15mm, clip=true]{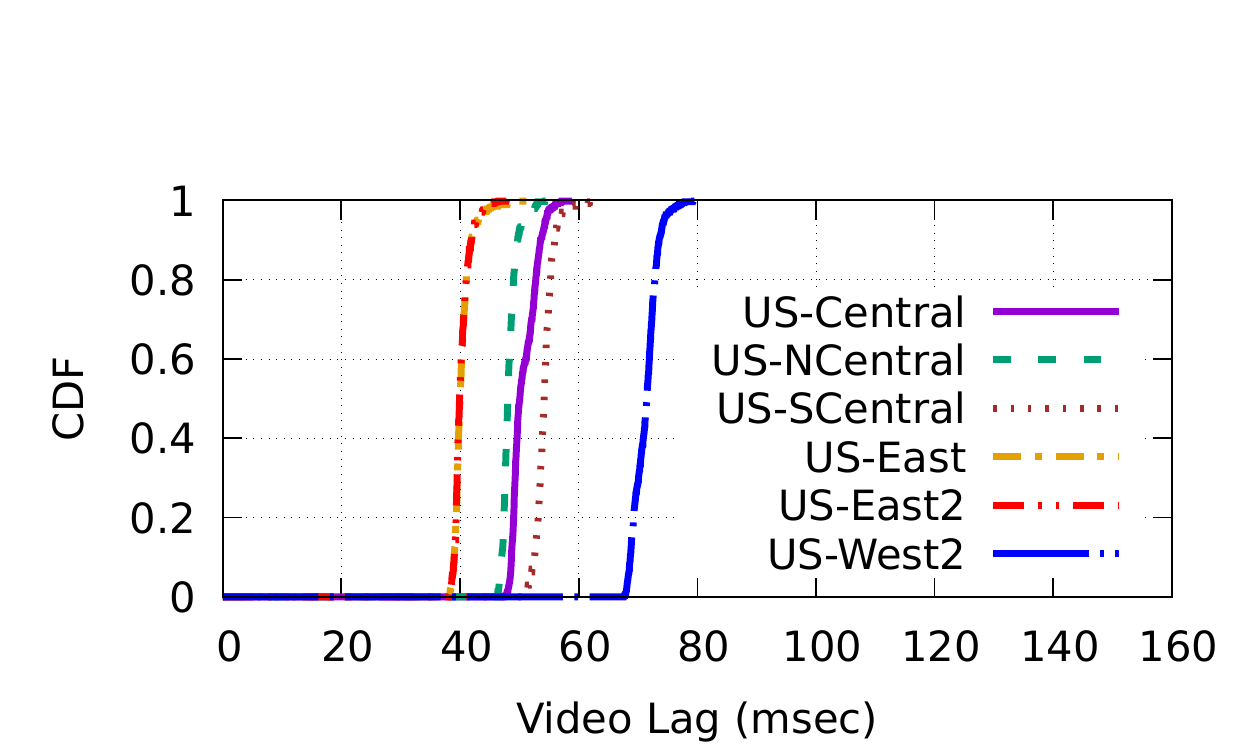}
\caption{Webex}
\label{fig:lag_uswest:webex}
\end{subfigure}
\begin{subfigure}{0.32\linewidth}
\centering
\includegraphics[width = 1.0\textwidth,trim = 0mm 0mm 0mm 15mm, clip=true]{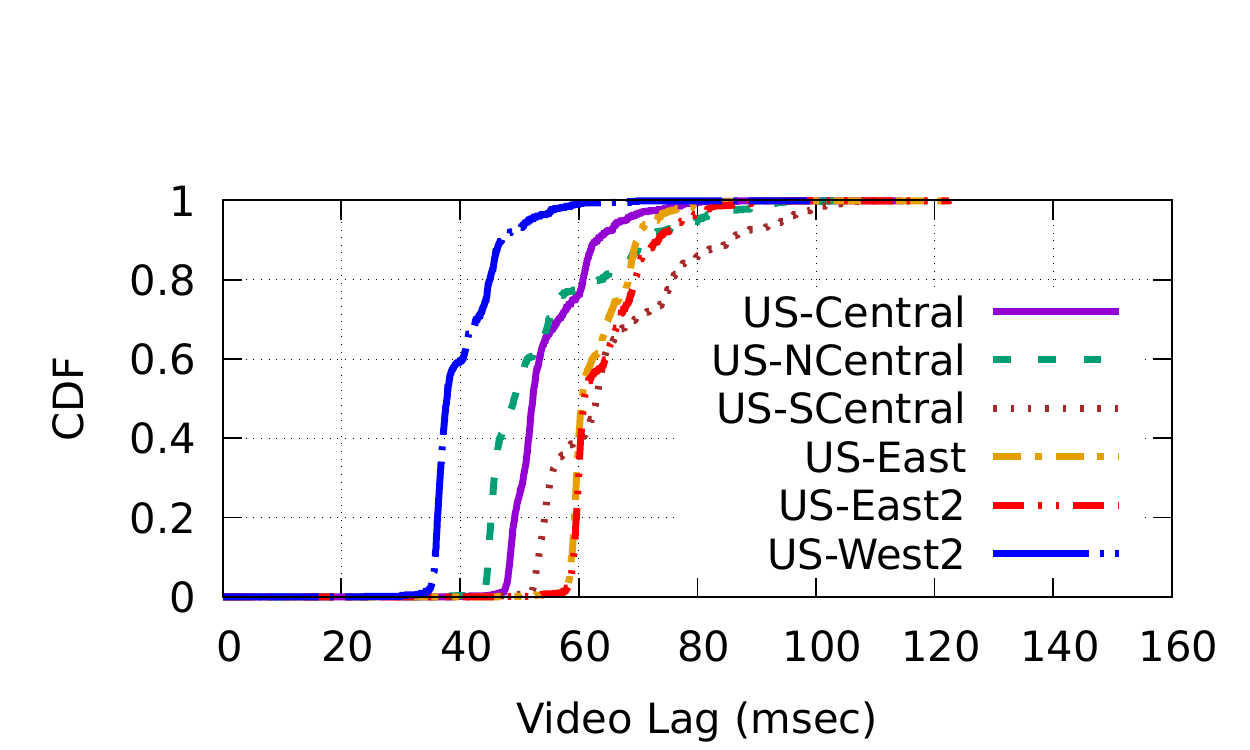}
\caption{Meet}
\label{fig:lag_uswest:goog}
\end{subfigure}
\vspace{-1ex}
\caption{CDF of streaming lag: meeting host in US-west.}
\label{fig:lag_uswest}
\vspace*{-2ex}
\end{figure*}

\begin{figure*}[ht]
\centering
\begin{subfigure}{0.32\linewidth}
\centering
\includegraphics[width = 1.0\textwidth,trim = 0mm 0mm 0mm 15mm, clip=true]{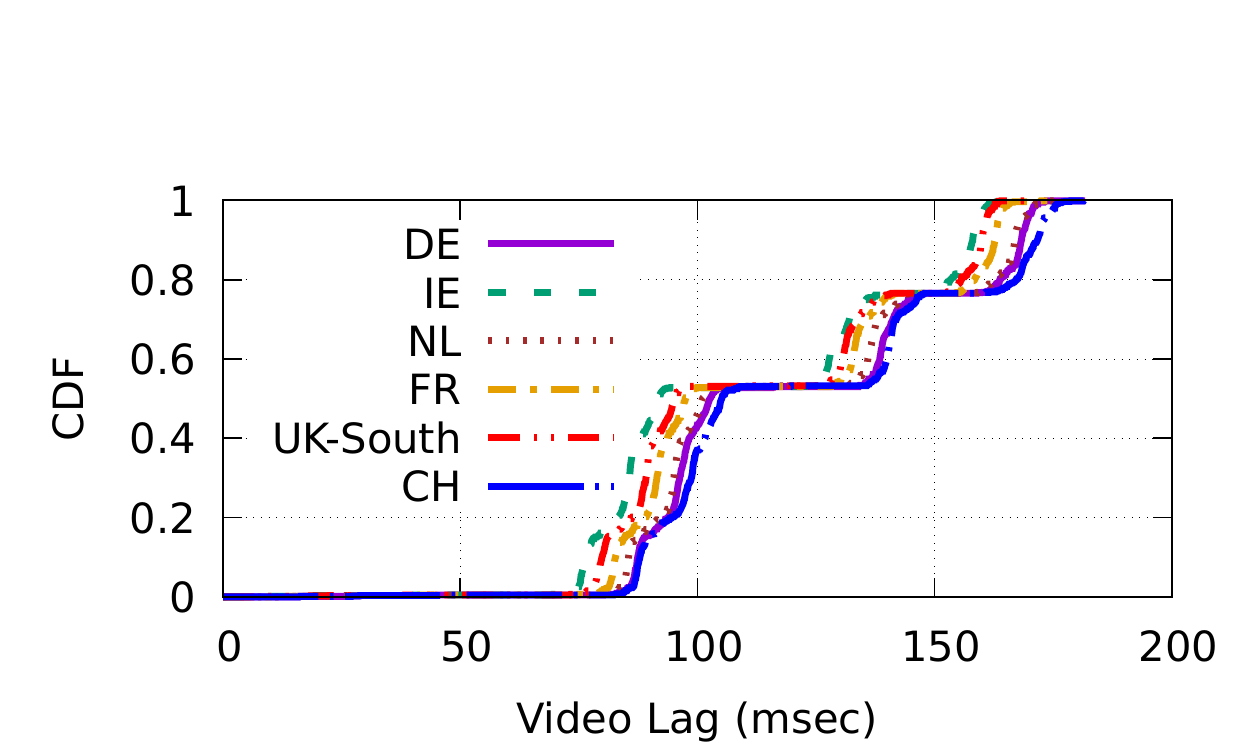}
\caption{Zoom}
\label{fig:lag_ukw:zoom}
\end{subfigure}
\begin{subfigure}{0.32\linewidth}
\centering
\includegraphics[width = 1.0\textwidth,trim = 0mm 0mm 0mm 15mm, clip=true]{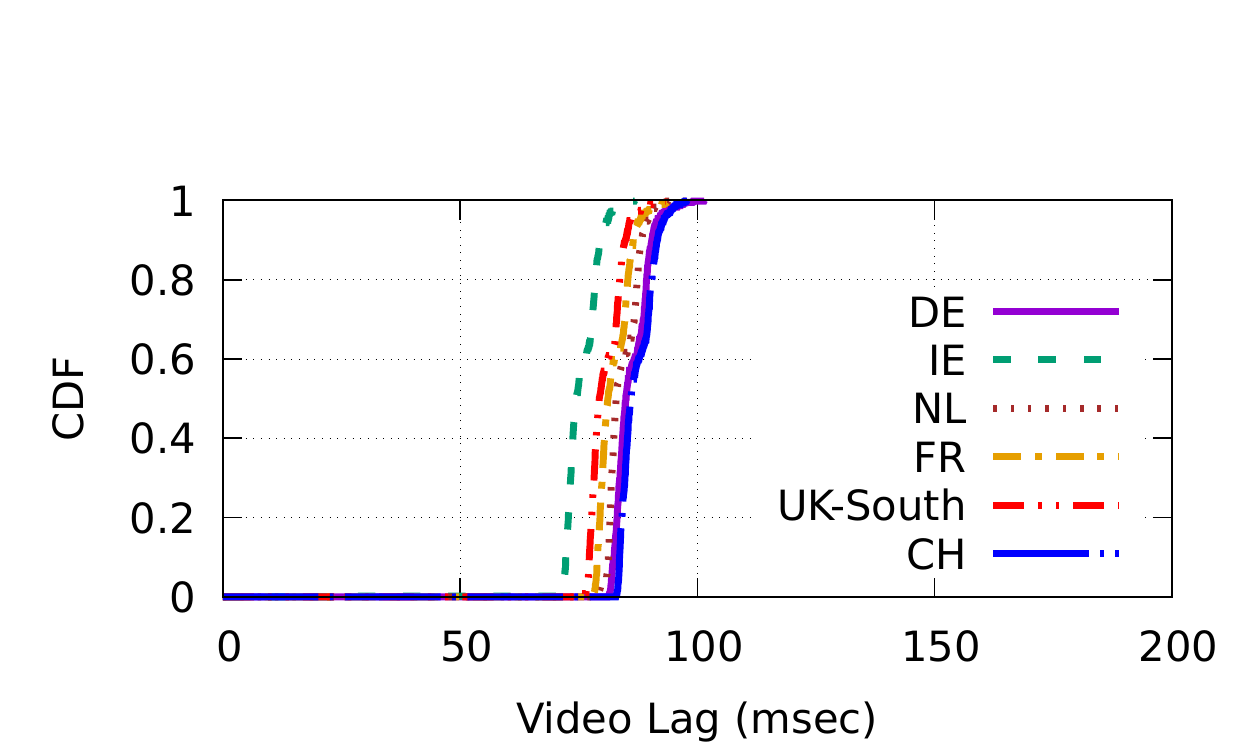}
\caption{Webex}
\label{fig:lag_ukw:webex}
\end{subfigure}
\begin{subfigure}{0.32\linewidth}
\centering
\includegraphics[width = 1.0\textwidth,trim = 0mm 0mm 0mm 15mm, clip=true]{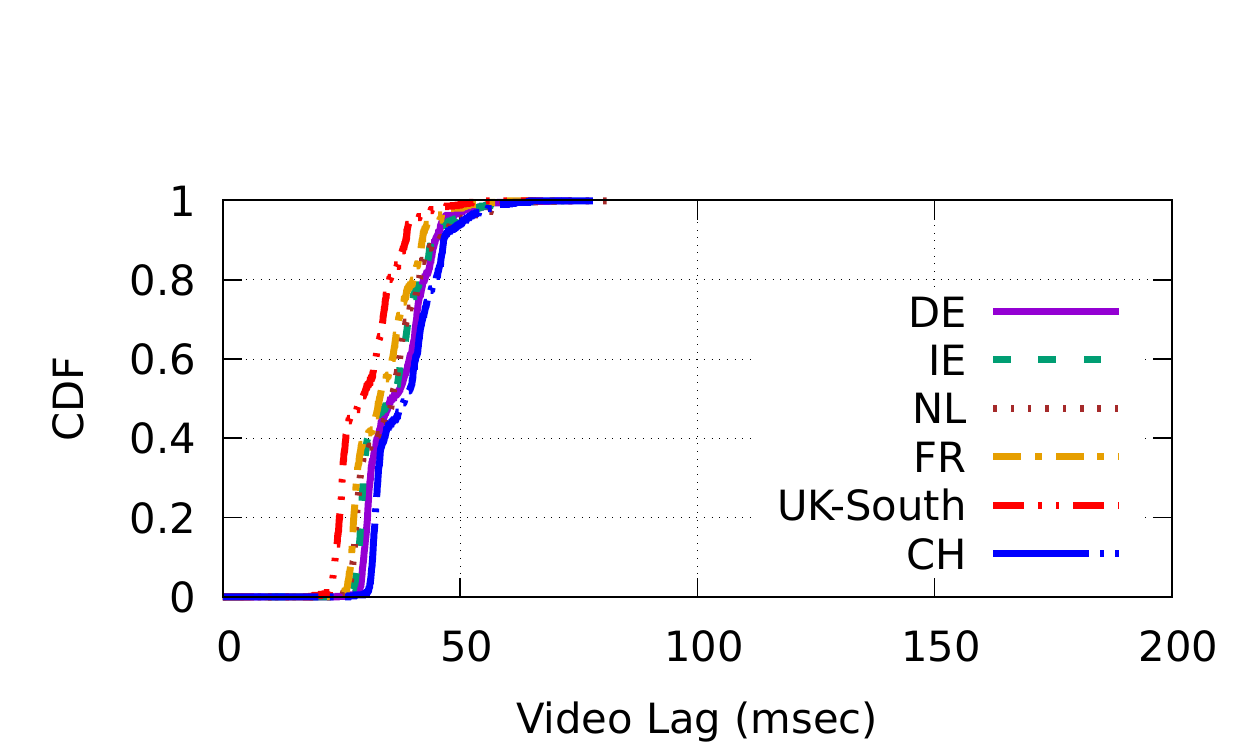}
\caption{Meet}
\label{fig:lag_ukw:goog}
\end{subfigure}
\vspace{-1ex}
\caption{CDF of streaming lag: meeting host in UK-west.}
\label{fig:lag_ukw}
\vspace*{-2ex}
\end{figure*}

\begin{figure*}[ht]
\centering
\begin{subfigure}{0.32\linewidth}
\centering
\includegraphics[width = 1.0\textwidth,trim = 0mm 0mm 0mm 15mm, clip=true]{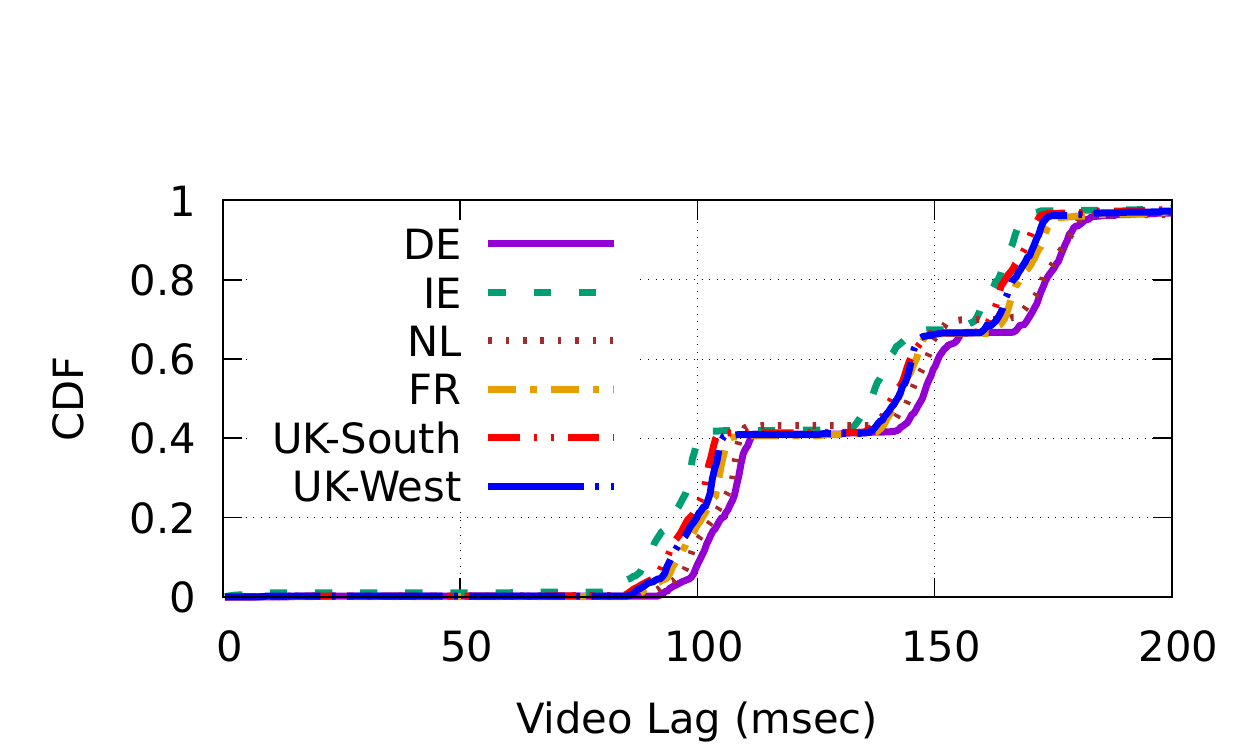}
\caption{Zoom}
\label{fig:lag_ch:zoom}
\end{subfigure}
\begin{subfigure}{0.32\linewidth}
\centering
\includegraphics[width = 1.0\textwidth,trim = 0mm 0mm 0mm 15mm, clip=true]{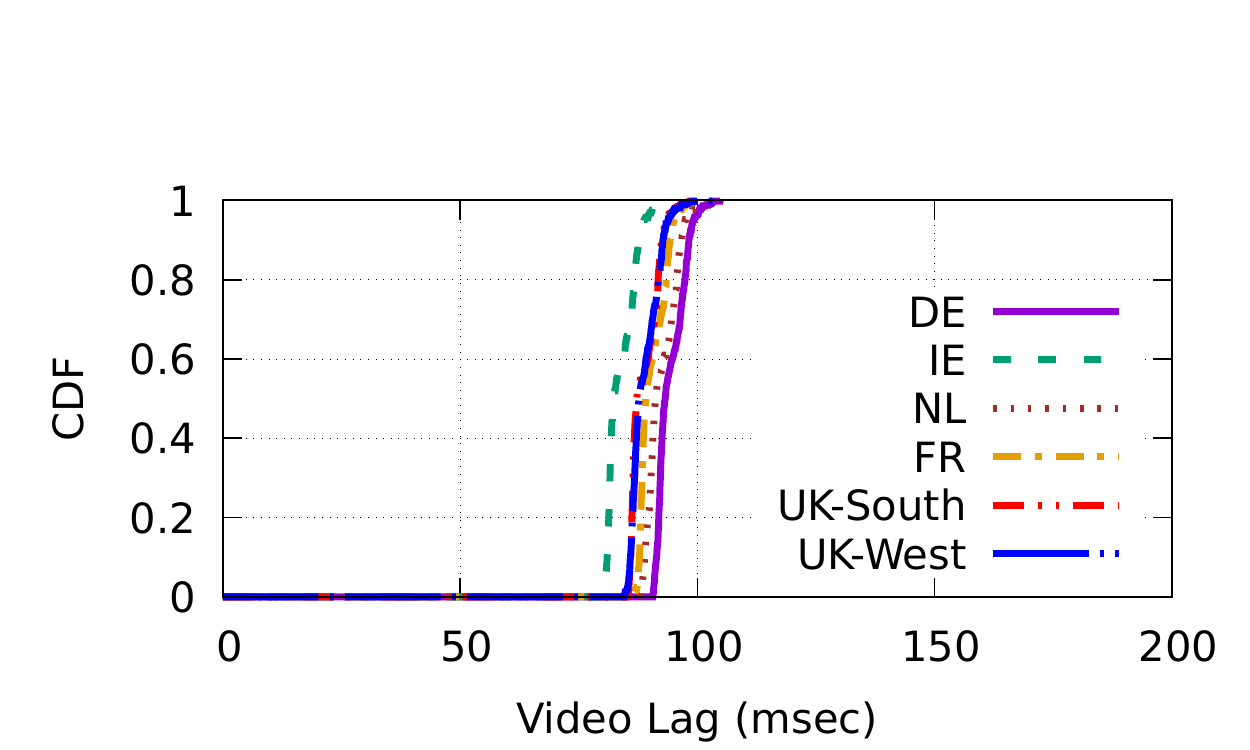}
\caption{Webex}
\label{fig:lag_ch:webex}
\end{subfigure}
\begin{subfigure}{0.32\linewidth}
\centering
\includegraphics[width = 1.0\textwidth,trim = 0mm 0mm 0mm 15mm, clip=true]{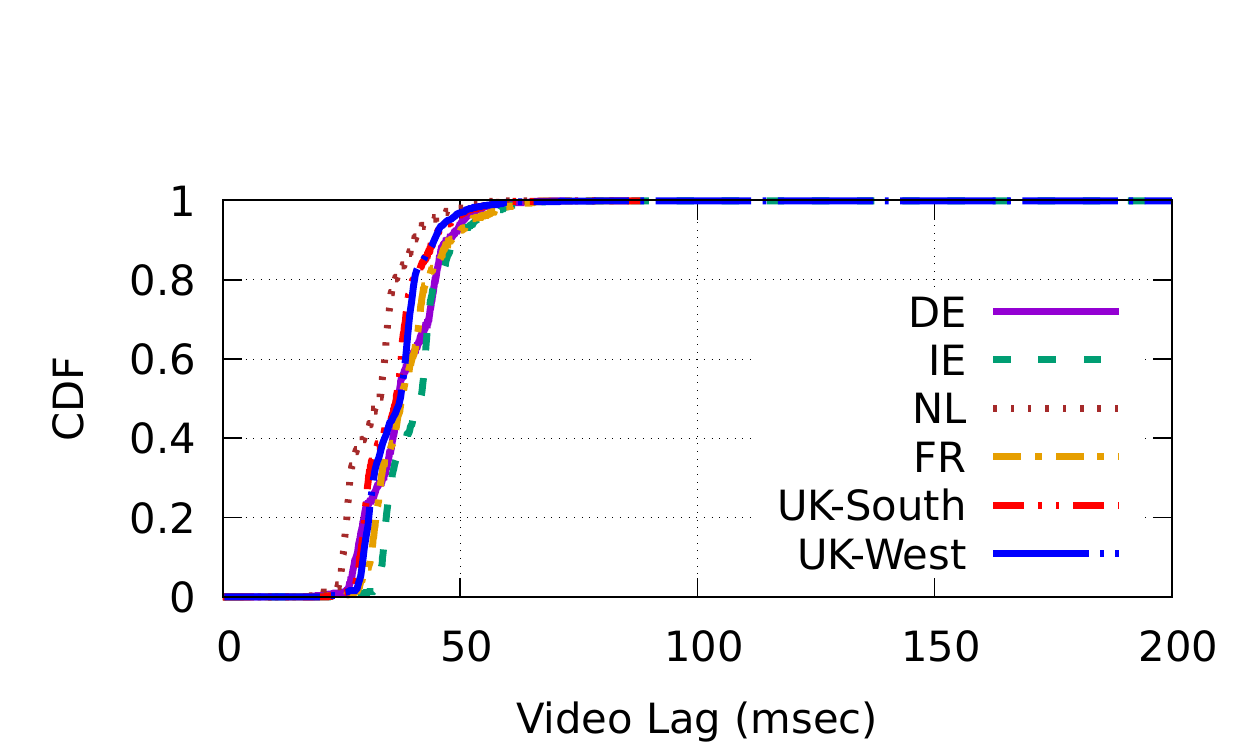}
\caption{Meet}
\label{fig:lag_ch:goog}
\end{subfigure}
\vspace{-1ex}
\caption{CDF of streaming lag: meeting host in Switzerland.}
\label{fig:lag_ch}
\vspace*{-2ex}
\end{figure*}

We observe that the multi-user sessions created in this experiment are all relayed via platform-operated service endpoints with a designated fixed port number (\texttt{UDP/8801} for Zoom, \texttt{UDP/9000} for Webex, and \texttt{UDP/19305} for Meet).\footnote{One exceptional case we observe is that, on Zoom, if there are only two users in a session, peer-to-peer streaming is activated, where they stream to each other \emph{directly} on an ephemeral port without going through an intermediary service endpoint.}  Fig.~\ref{fig:topo} compares the three videoconferencing systems in terms of how their clients interact with the service endpoints for content streaming, which we discover from their traffic traces.  On Zoom and Webex, a single service endpoint is designated for each meeting session, and all meeting participants send or receive streaming data via this endpoint. On Meet, each client connects to a separate (geographically close-by) endpoint, and meeting sessions are relayed among these multiple distinct endpoints.


\begin{figure*}[ht]
\centering
\begin{subfigure}{0.32\linewidth}
\centering
\includegraphics[width = 1.0\textwidth,trim = 0mm 5mm 0mm 20mm, clip=true]{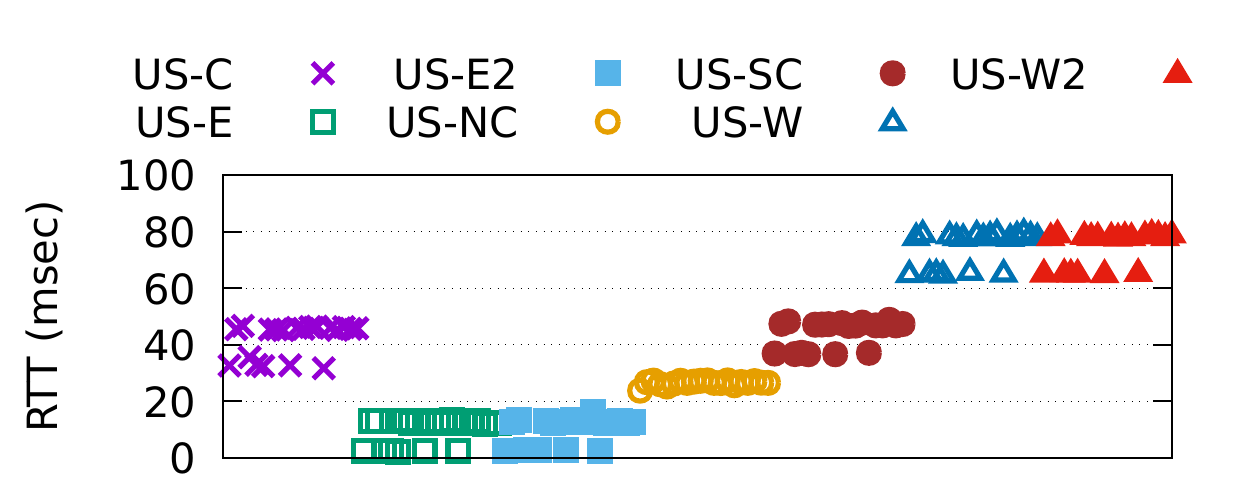}
\caption{Zoom}
\label{fig:rtt_useast:zoom}
\end{subfigure}
\begin{subfigure}{0.32\linewidth}
\centering
\includegraphics[width = 1.0\textwidth,trim = 0mm 5mm 0mm 20mm, clip=true]{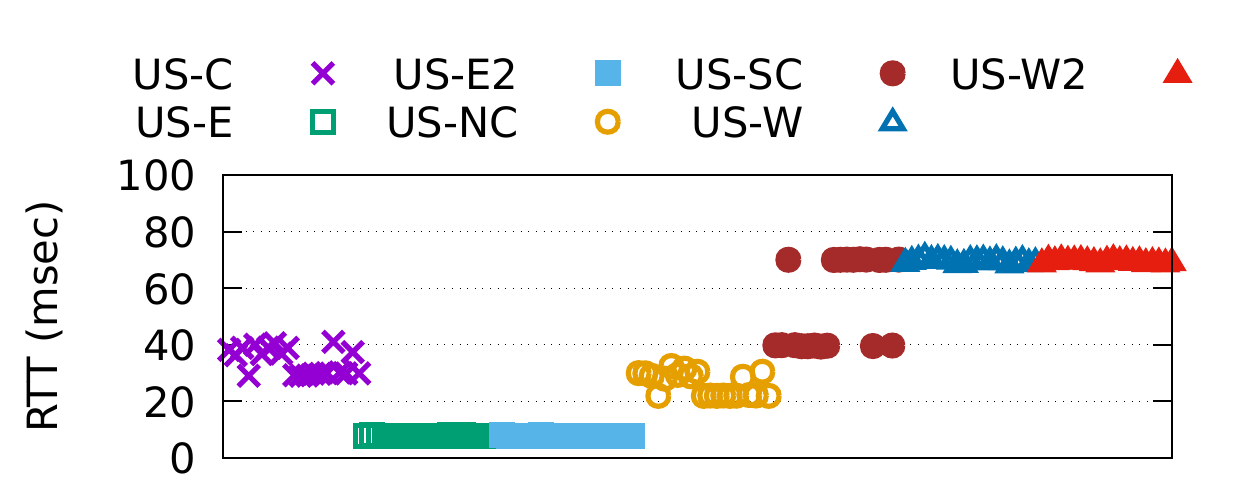}
\caption{Webex}
\label{fig:rtt_useast:webex}
\end{subfigure}
\begin{subfigure}{0.32\linewidth}
\centering
\includegraphics[width = 1.0\textwidth,trim = 0mm 5mm 0mm 20mm, clip=true]{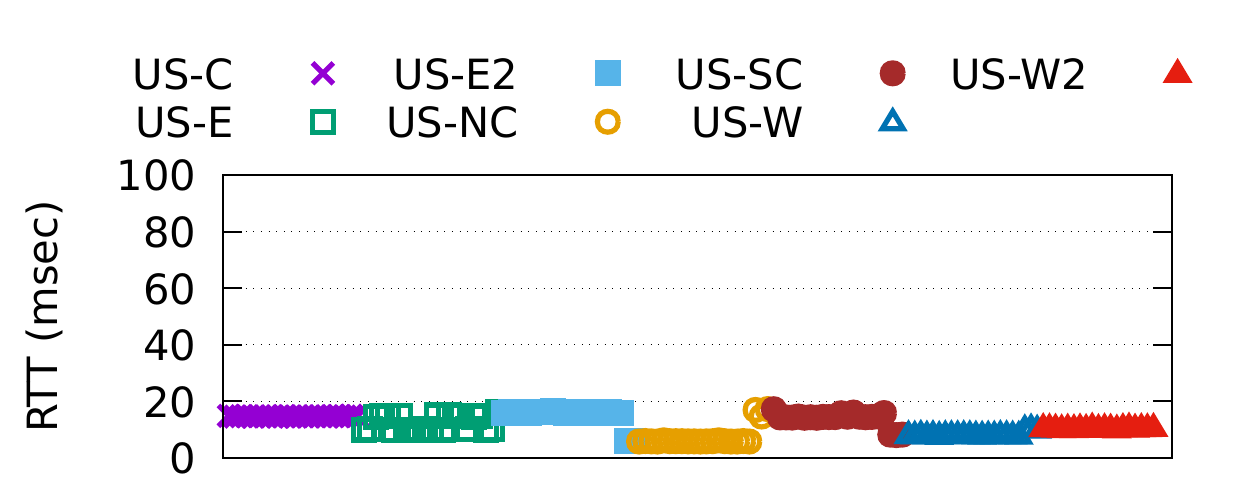}
\caption{Meet}
\label{fig:rtt_useast:goog}
\end{subfigure}
\vspace{-2ex}
\caption{Service proximity: meeting host in US-east.}
\vspace*{-1ex}
\label{fig:rtt_useast}
\end{figure*}

\begin{figure*}[ht]
\centering
\begin{subfigure}{0.32\linewidth}
\centering
\includegraphics[width = 1.0\textwidth,trim = 0mm 5mm 0mm 20mm, clip=true]{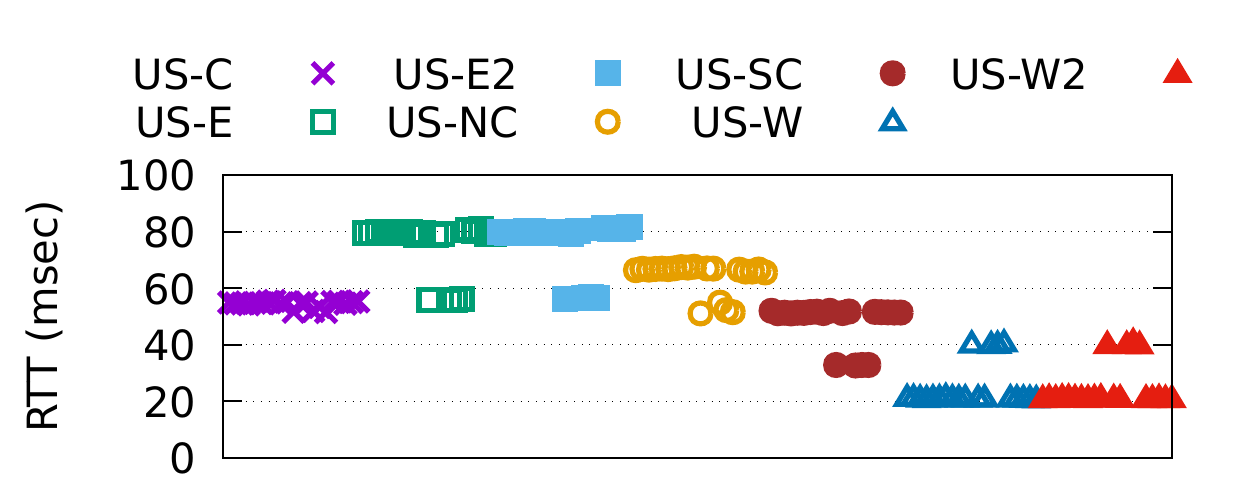}
\caption{Zoom}
\label{fig:rtt_uswest:zoom}
\end{subfigure}
\begin{subfigure}{0.32\linewidth}
\centering
\includegraphics[width = 1.0\textwidth,trim = 0mm 5mm 0mm 20mm, clip=true]{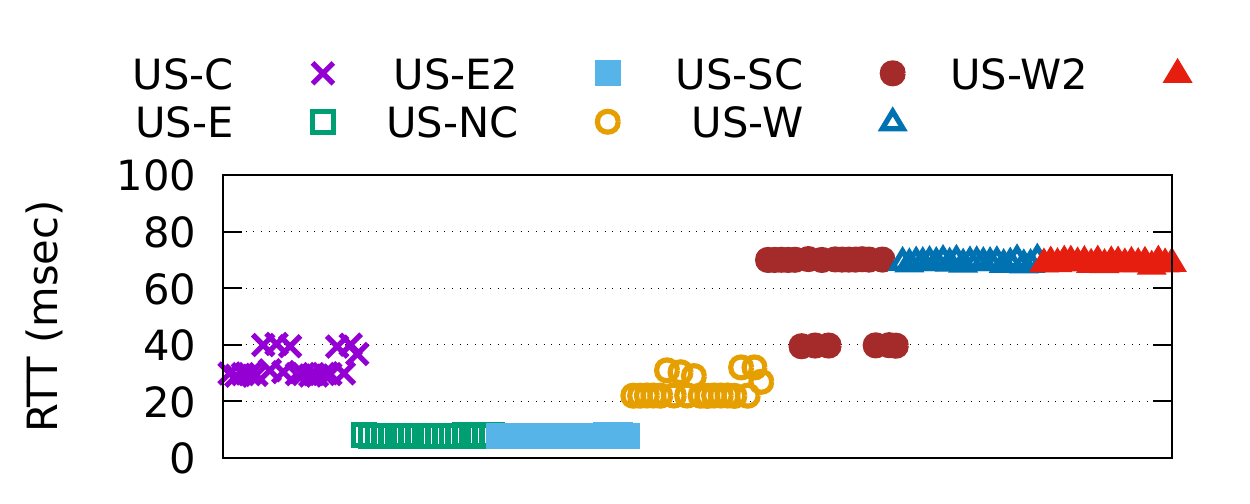}
\caption{Webex}
\label{fig:rtt_uswest:webex}
\end{subfigure}
\begin{subfigure}{0.32\linewidth}
\centering
\includegraphics[width = 1.0\textwidth,trim = 0mm 5mm 0mm 20mm, clip=true]{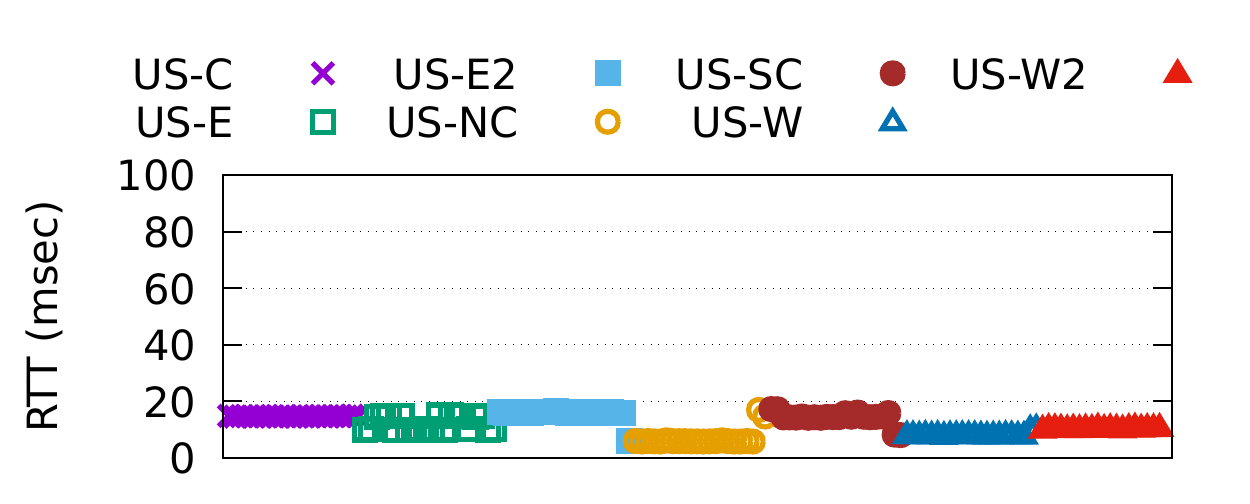}
\caption{Meet}
\label{fig:rtt_uswest:goog}
\end{subfigure}
\vspace{-1ex}
\caption{Service proximity: meeting host in US-west.}
\vspace{-1ex}
\label{fig:rtt_uswest}
\end{figure*}

\begin{figure*}[ht]
\centering
\begin{subfigure}{0.32\linewidth}
\centering
\includegraphics[width = 1.0\textwidth,trim = 0mm 5mm 0mm 20mm, clip=true]{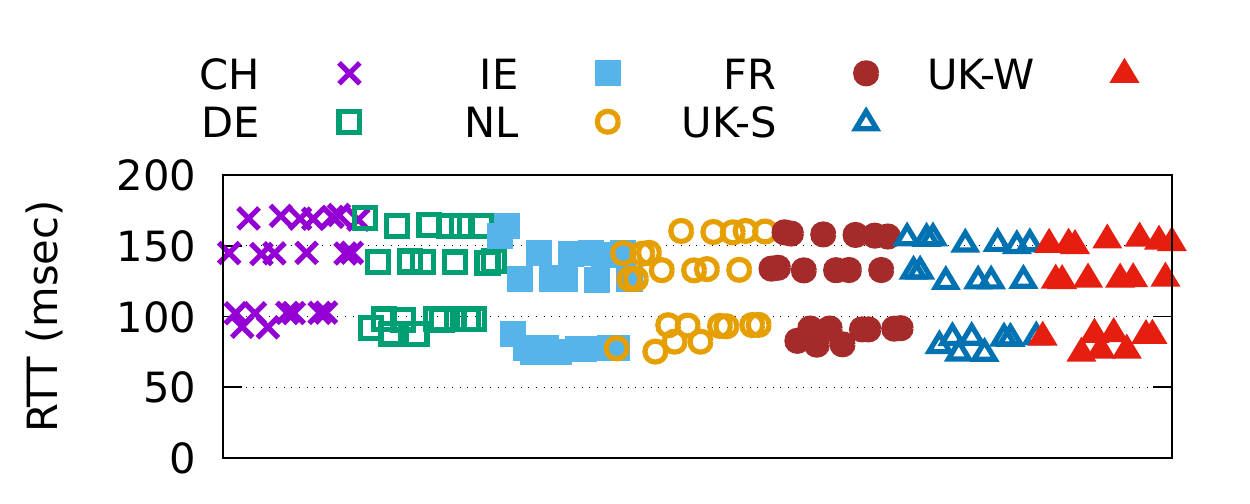}
\caption{Zoom}
\label{fig:rtt_ukw:zoom}
\end{subfigure}
\begin{subfigure}{0.32\linewidth}
\centering
\includegraphics[width = 1.0\textwidth,trim = 0mm 5mm 0mm 20mm, clip=true]{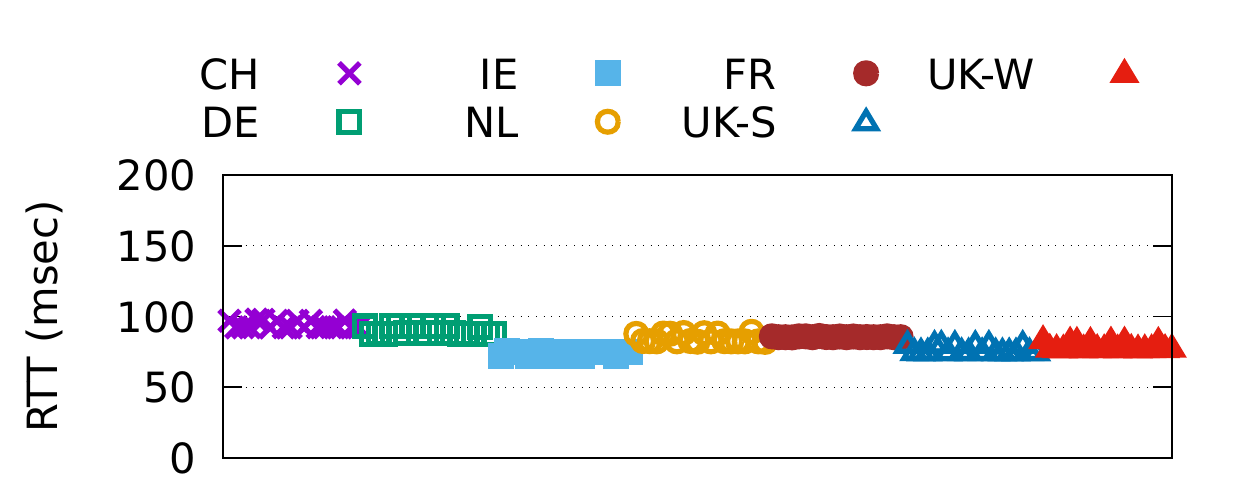}
\caption{Webex}
\label{fig:rtt_ukw:webex}
\end{subfigure}
\begin{subfigure}{0.32\linewidth}
\centering
\includegraphics[width = 1.0\textwidth,trim = 0mm 5mm 0mm 20mm, clip=true]{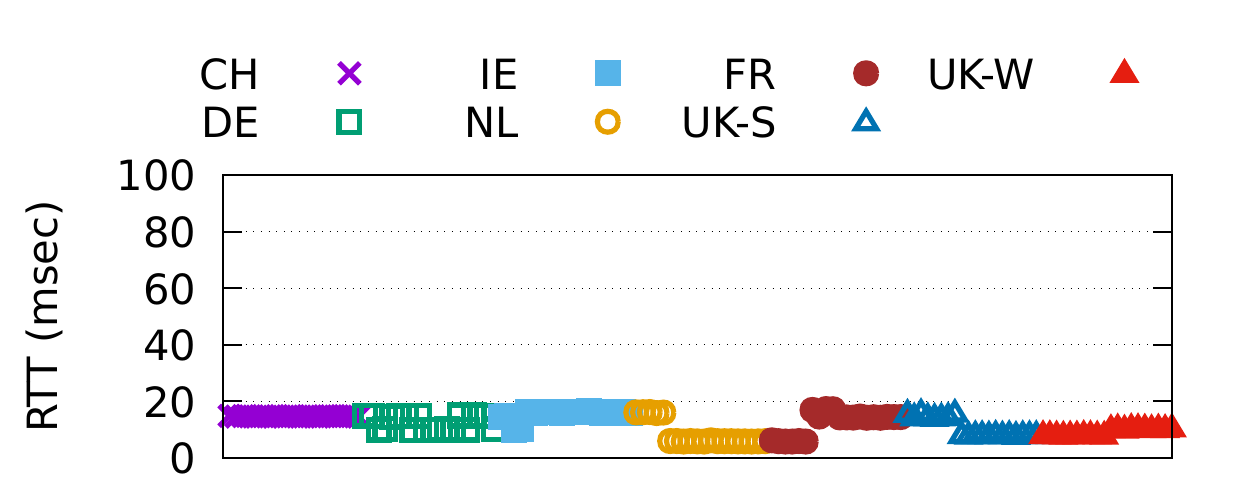}
\caption{Meet}
\label{fig:rtt_ukw:goog}
\end{subfigure}
\vspace{-1ex}
\caption{Service proximity: meeting host in UK-west.}
\vspace{-1ex}
\label{fig:rtt_ukw}
\end{figure*}

\begin{figure*}[ht]
\centering
\begin{subfigure}{0.32\linewidth}
\centering
\includegraphics[width = 1.0\textwidth,trim = 0mm 5mm 0mm 20mm, clip=true]{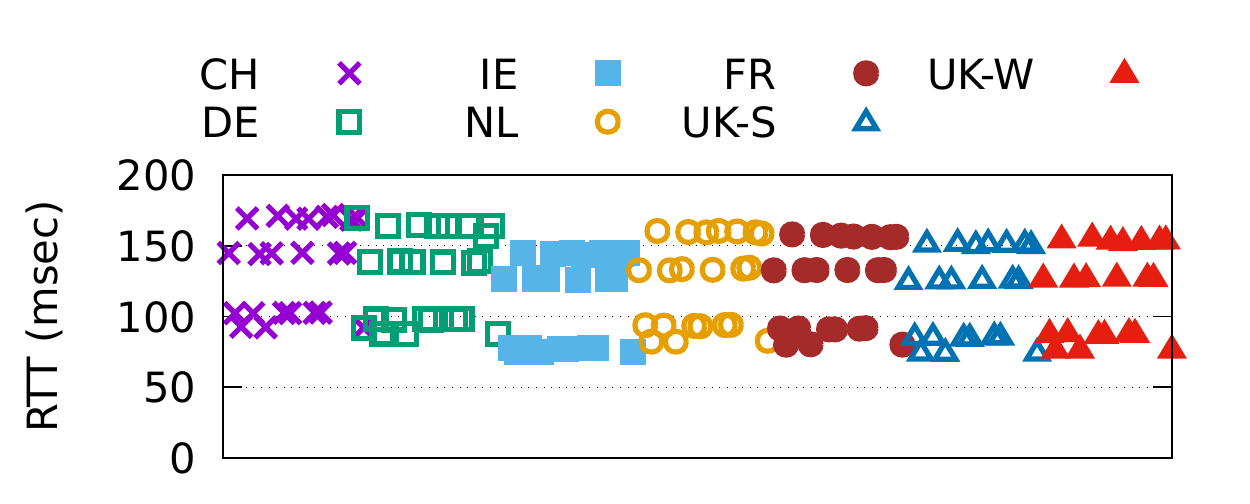}
\caption{Zoom}
\label{fig:rtt_ch:zoom}
\end{subfigure}
\begin{subfigure}{0.32\linewidth}
\centering
\includegraphics[width = 1.0\textwidth,trim = 0mm 5mm 0mm 20mm, clip=true]{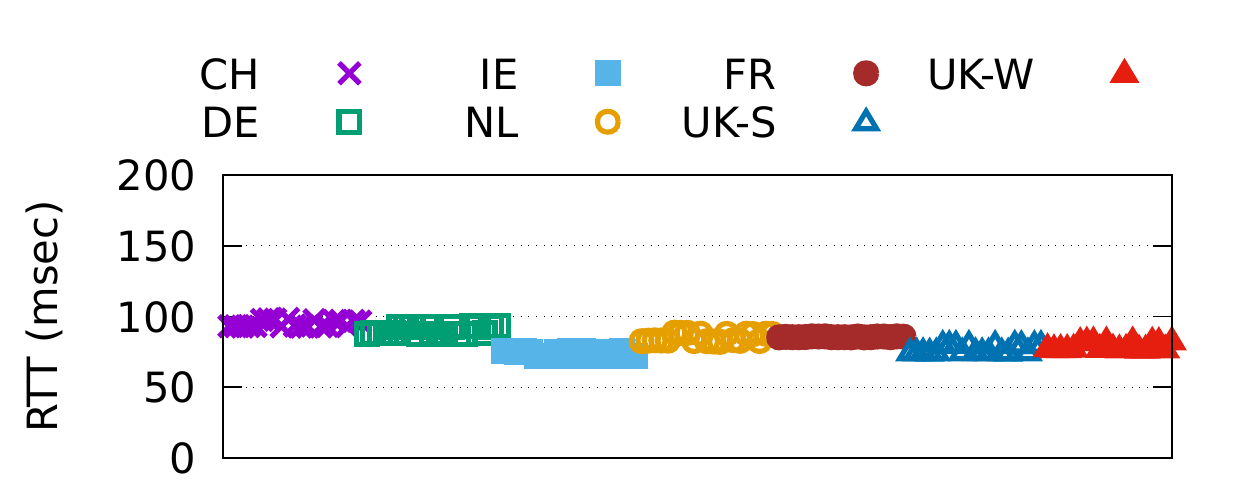}
\caption{Webex}
\label{fig:rtt_ch:webex}
\end{subfigure}
\begin{subfigure}{0.32\linewidth}
\centering
\includegraphics[width = 1.0\textwidth,trim = 0mm 5mm 0mm 20mm, clip=true]{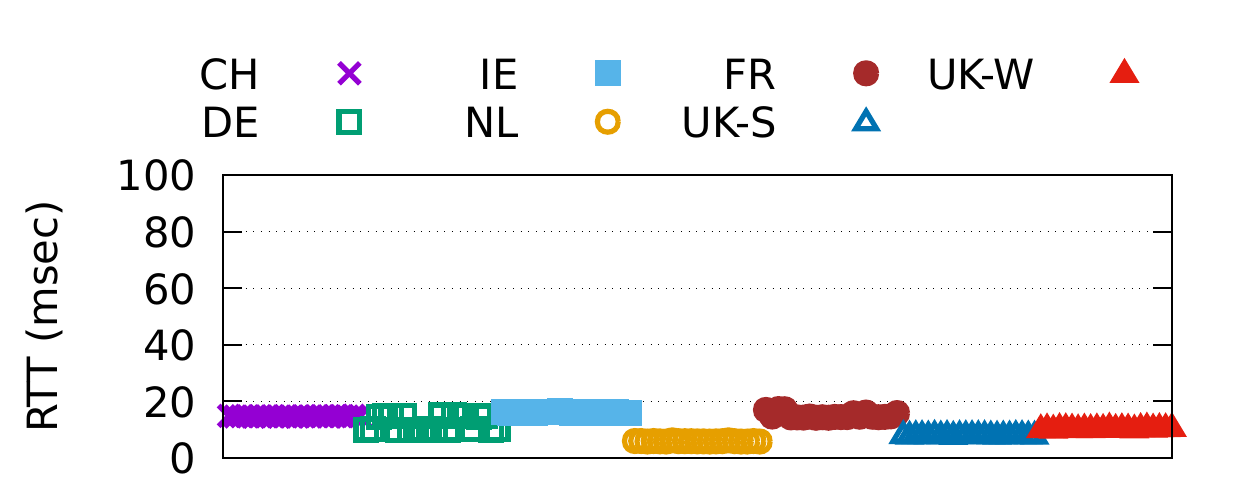}
\caption{Meet}
\label{fig:rtt_ch:goog}
\end{subfigure}
\vspace{-1ex}
\caption{Service proximity: meeting host in Switzerland.}
\vspace{-1ex}
\label{fig:rtt_ch}
\end{figure*}

The number of distinct service endpoints encountered by a client varies greatly across different platforms. For example, out of 20 videoconferencing sessions, a client on Zoom, Webex and Meet encounters, on average, 20, 19.5 and 1.8 endpoints, respectively.  On Zoom and Webex, service endpoints almost always change (with different IP addresses) across different sessions, while, on Meet, a client tends to stick with one or two endpoints across sessions.

Figs.~\ref{fig:lag_useast}--\ref{fig:lag_ch} plot the CDFs of streaming lag experienced by clients in four different scenarios.  In Figs.~\ref{fig:lag_useast} and \ref{fig:lag_uswest}, we consider videoconferencing sessions among seven US-based clients (including a meeting host), where the host is located in either US-east or US-west. In Figs.~\ref{fig:lag_ukw}~and~\ref{fig:lag_ch}, similarly we set up videoconferencing sessions among seven clients in Europe, with a meeting host in either UK or Switzerland.  We make the following observations from the results.

\subsubsection{US-based Videoconferencing}
When sessions are created from US-east (Fig.~\ref{fig:lag_useast}), across all three platforms, streaming lag experienced by clients increases as they are further away from US-east, with the US-west clients experiencing the most lags (about \SI{30}{\ms} higher than the US-east client). This implies that streaming is relayed via the servers in US-east, where the meeting host resides.  This is in fact confirmed by Fig.~\ref{fig:rtt_useast}, where we plot RTTs between clients and service endpoints they are connected to. In the figure, RTTs measured by different clients are indicated with distinct dots.  Each dot represents an average RTT (over 100 measurements) in a particular session.  On Zoom and Webex, RTTs measured by US-east users are much lower than those by US-west users. On Meet, RTTs are uniform across clients due to its distributed service endpoint architecture as shown in Fig.~\ref{fig:topo}.

When a meeting host is in US-west (Fig.~\ref{fig:lag_uswest}), geographic locality plays a similar role with Zoom and Meet, where the most far-away clients in US-east experience the worst lags.  In case of Webex, however, the worst streaming lag is actually experienced by another user in US-west.  According to the RTTs collected in this scenario for Webex (Fig.~\ref{fig:rtt_uswest:webex}), its service endpoints seem to be provisioned on the \emph{east}-side of the US even when sessions are created in US-west, causing the streaming between US-west users to be detoured via US-east.  Due to the geographically-skewed infrastructure, the lag distributions for US-west-based sessions are simply shifted by \SI{30}{\ms} from the US-east-based counterparts (Figs.~\ref{fig:lag_useast:webex} and~\ref{fig:lag_uswest:webex}).
One unexpected observation is that Meet sessions exhibit the worst lag despite having the lowest RTTs.  This might be due to the fact that Meet sessions are relayed via \emph{multiple} service endpoints, unlike Zoom and Webex (Fig.~\ref{fig:topo}).  In addition, although Meet's videoconferencing infrastructure appears to be distributed over wider locations (with lower RTTs), the total aggregate server capacity at each location may be smaller, hence leading to more load variation.

\begin{figure*}[ht]
\centering
\begin{subfigure}{0.32\linewidth}
\centering
\includegraphics[width = 1.0\textwidth,trim = 5mm 1mm 5mm 5mm, clip=true]{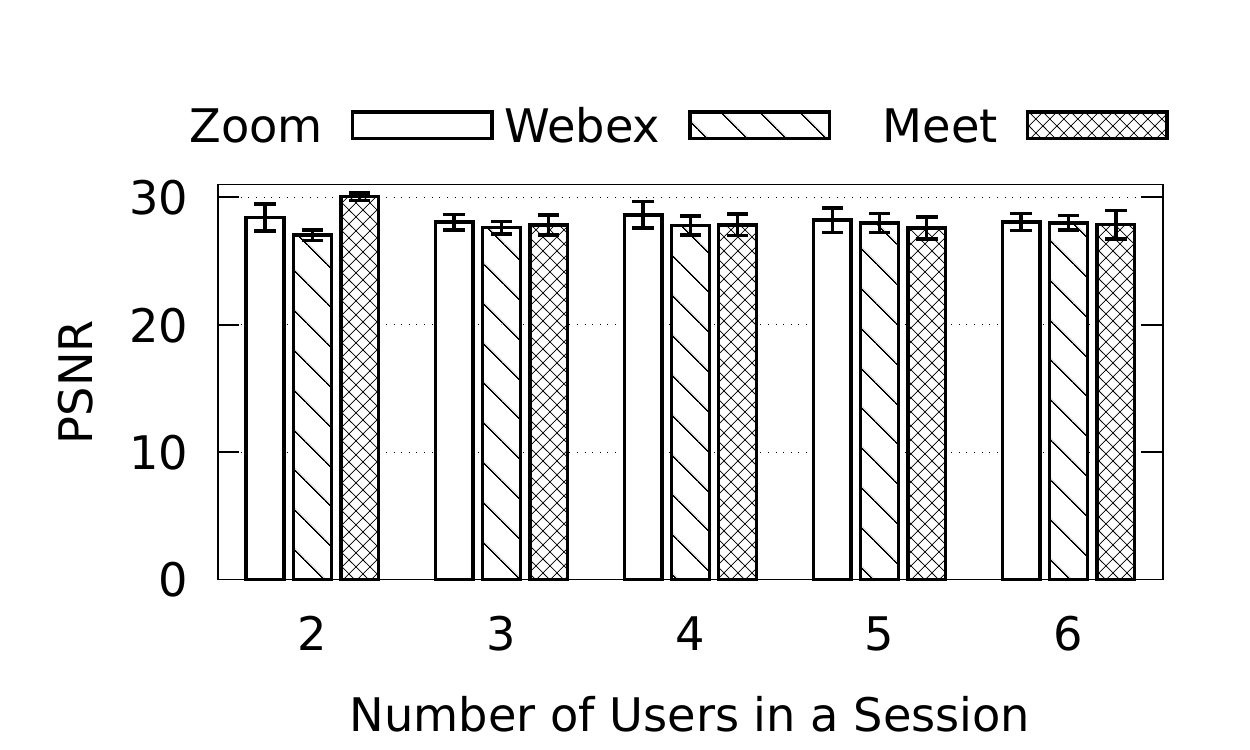}
\caption{Low motion: PSNR}
\label{fig:scale_us:lm:psnr}
\end{subfigure}
\begin{subfigure}{0.32\linewidth}
\centering
\includegraphics[width = 1.0\textwidth,trim = 5mm 1mm 5mm 5mm, clip=true]{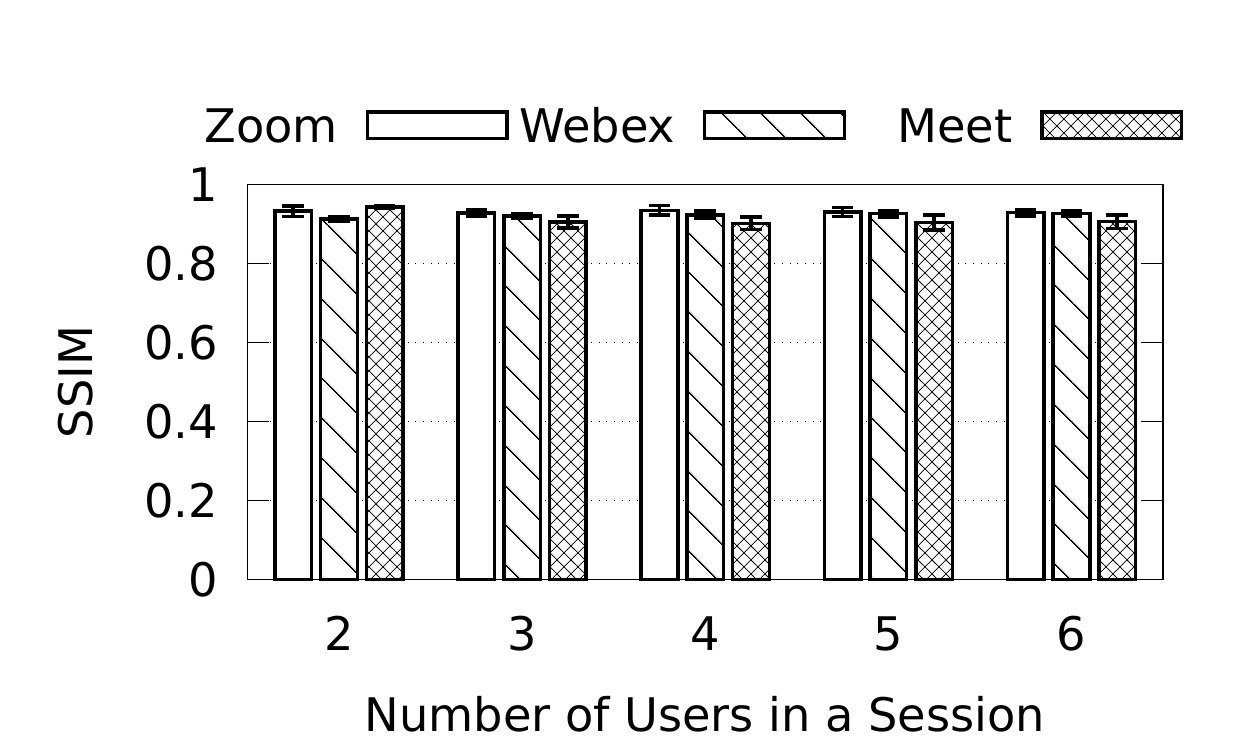}
\caption{Low motion: SSIM}
\label{fig:scale_us:lm:ssim}
\end{subfigure}
\begin{subfigure}{0.32\linewidth}
\centering
\includegraphics[width = 1.0\textwidth,trim = 5mm 1mm 5mm 5mm, clip=true]{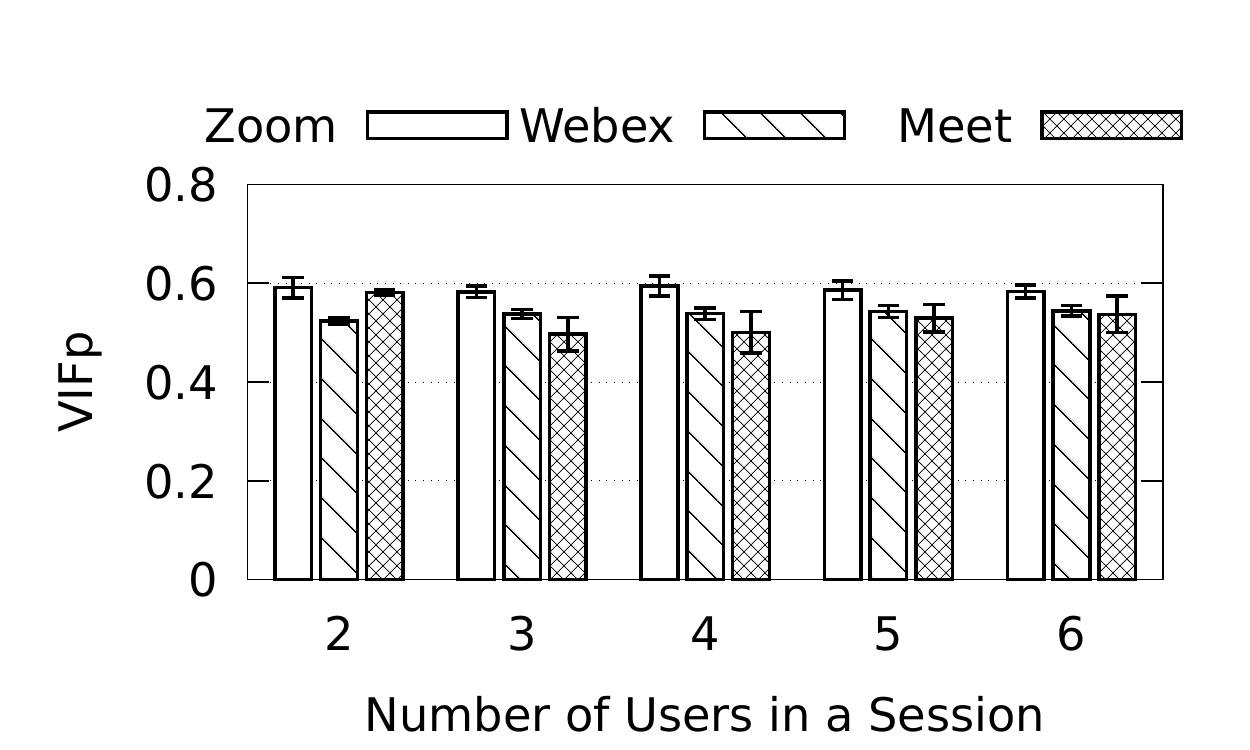}
\caption{Low motion: VIFp}
\label{fig:scale_us:lm:vifp}
\end{subfigure}
\begin{subfigure}{0.32\linewidth}
\centering
\includegraphics[width = 1.0\textwidth,trim = 5mm 1mm 5mm 5mm, clip=true]{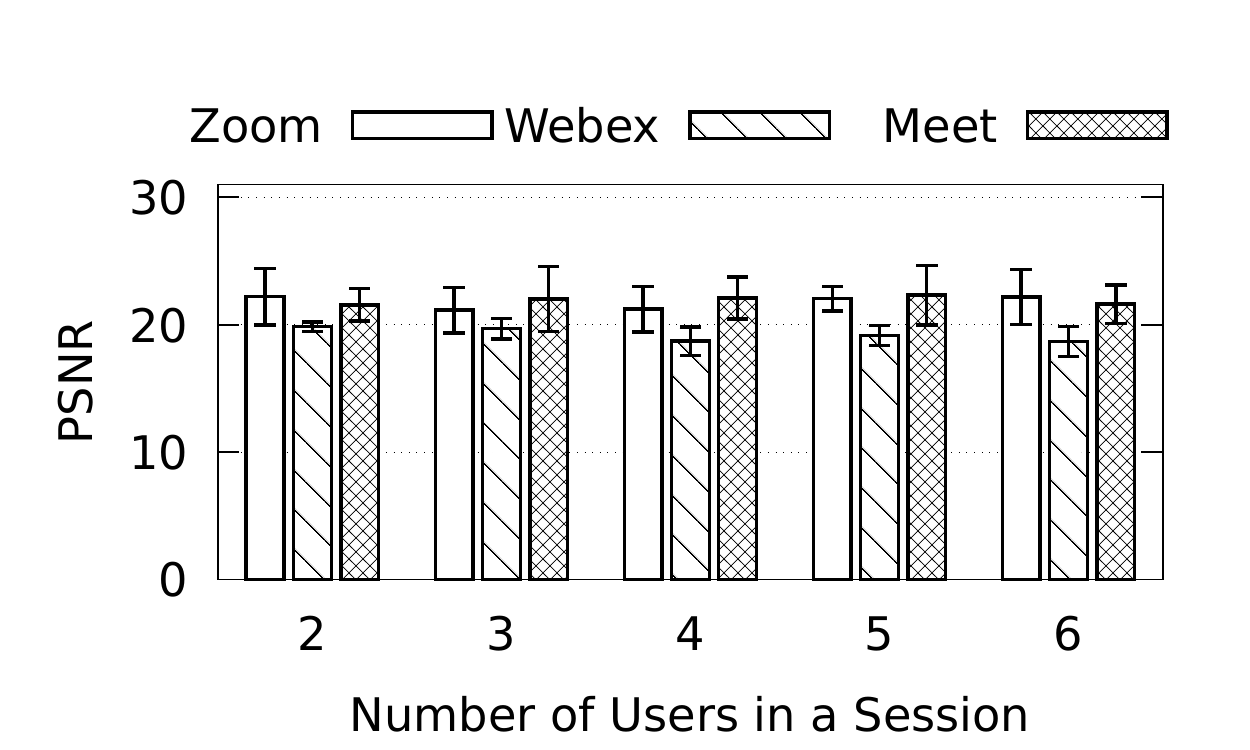}
\caption{High motion: PSNR}
\label{fig:scale_us:hm:psnr}
\end{subfigure}
\begin{subfigure}{0.32\linewidth}
\centering
\includegraphics[width = 1.0\textwidth,trim = 5mm 1mm 5mm 5mm, clip=true]{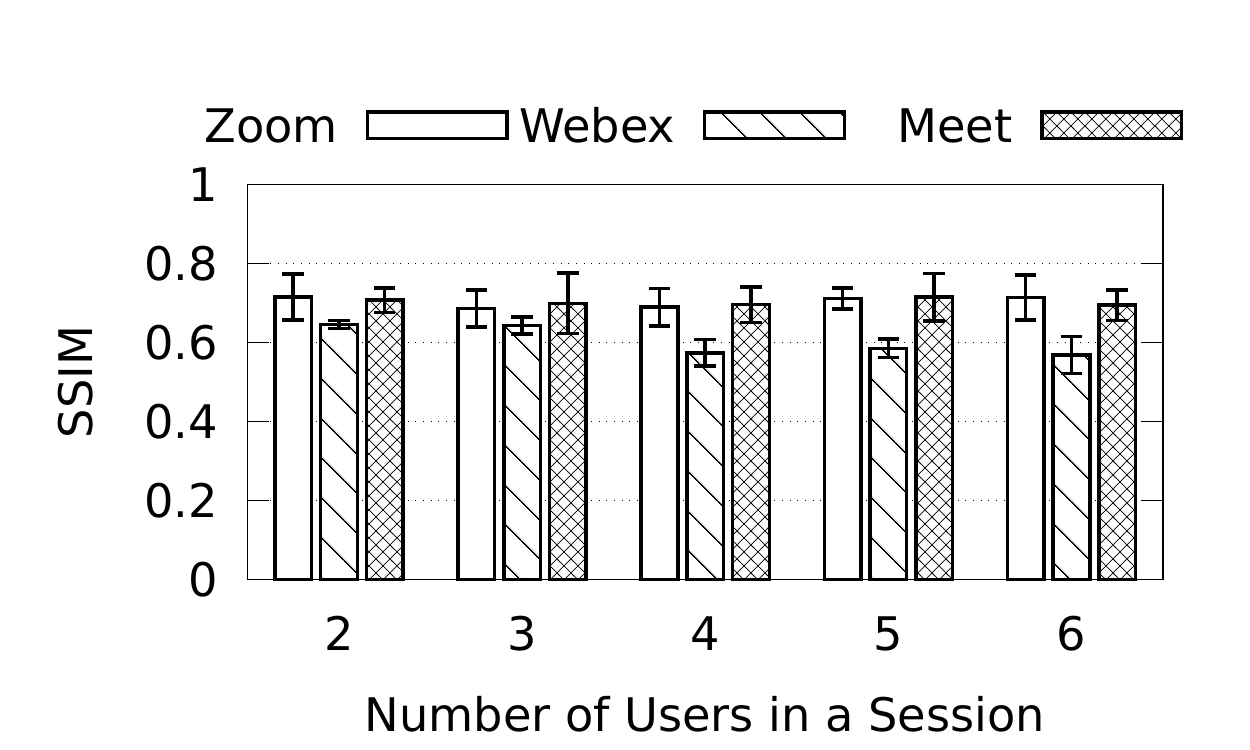}
\caption{High motion: SSIM}
\label{fig:scale_us:hm:ssim}
\end{subfigure}
\begin{subfigure}{0.32\linewidth}
\centering
\includegraphics[width = 1.0\textwidth,trim = 5mm 1mm 5mm 5mm, clip=true]{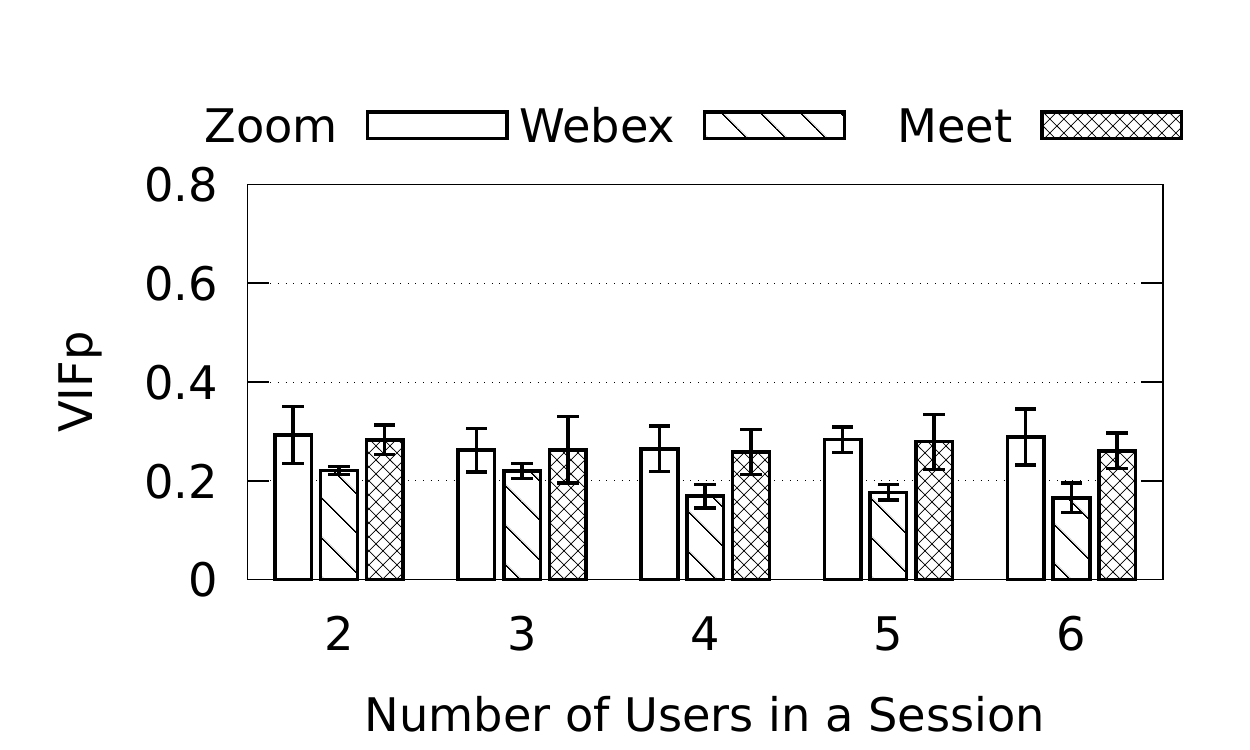}
\caption{High motion: VIFp}
\label{fig:scale_us:hm:vifp}
\end{subfigure}
\caption{Video QoE metrics comparison (US).}
\label{fig:scale_us}
\end{figure*}

\pagebreak
\subsubsection{Non-US-based videoconferencing} According to Figs.~\ref{fig:lag_ukw}--\ref{fig:lag_ch}, when sessions are set up among clients in Europe, Zoom/Webex clients experience much higher lags than Meet users. When compared to the Zoom/Webex sessions created in US-east, clients in Europe experience \SIrange[range-phrase=--,range-units=single]{55}{75}{\ms} and \SIrange[range-phrase=--,range-units=single]{45}{65}{\ms} higher median lags on Zoom and Webex, respectively.  The reported RTTs (Figs.~\ref{fig:rtt_ukw} and \ref{fig:rtt_ch}) show that the clients closer to the east-coast of US (\eg UK and Ireland) have lower RTTs than those located further into central Europe (\eg Germany and Switzerland). These observations suggest that the service infrastructures used are located somewhere in US.\footnote{We are not able to pinpoint the locations of the infrastructures because \texttt{traceroute}/\texttt{tcptraceroute}-probings are all blocked.}

\begin{figure}[t]
\centering
\vspace{1ex}
\includegraphics[width = 0.6\linewidth,trim = 0mm 0mm 0mm 0mm, clip=true]{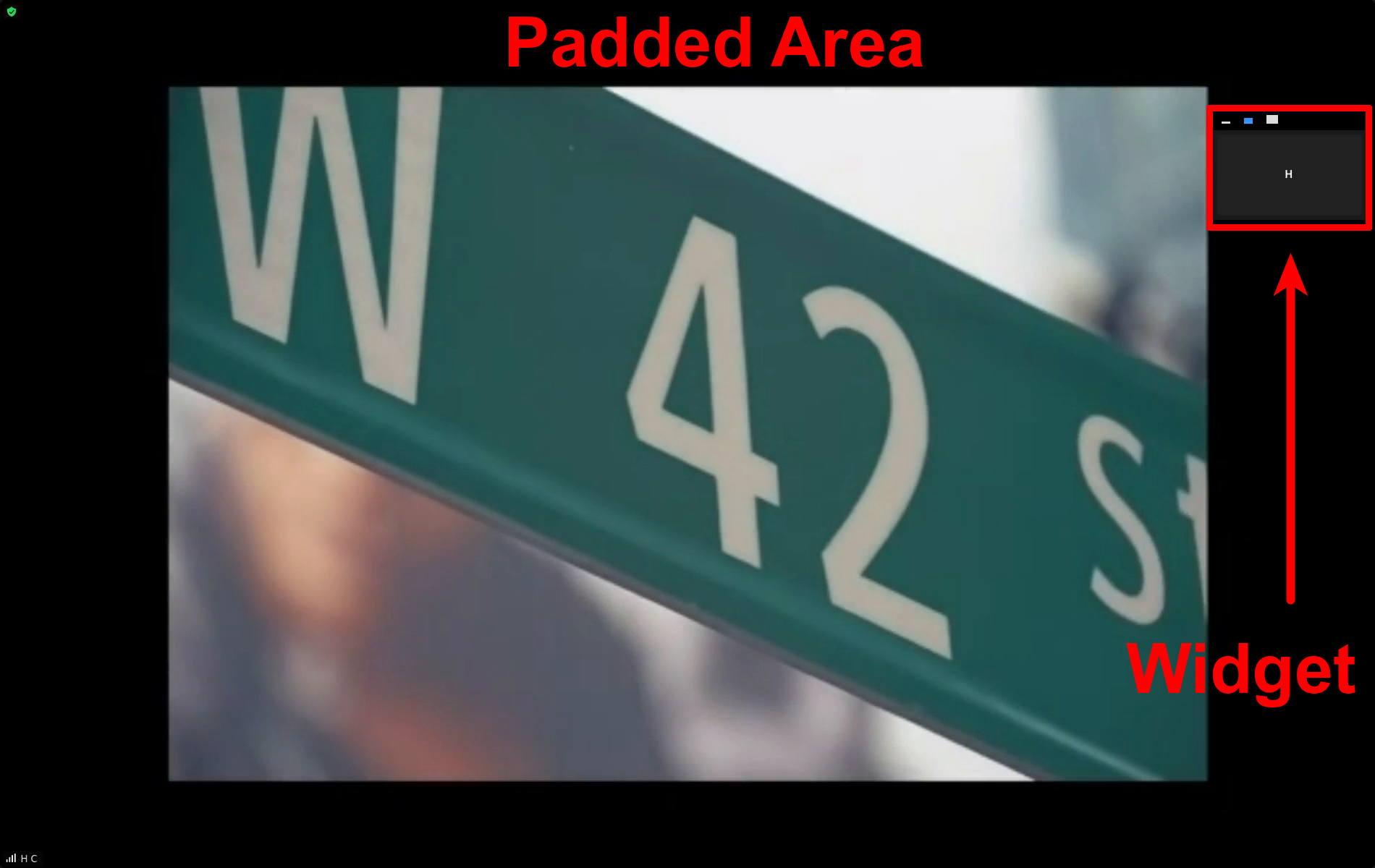}
\caption{Video screen with padding.}
\label{fig:padding}
\vspace{-3ex}
\end{figure}

Comparing Zoom and Webex, one can see that RTTs to service endpoints on Zoom vary much more widely across different sessions.  In fact, as shown in Figs.~\ref{fig:rtt_ukw:zoom} and \ref{fig:rtt_ch:zoom}, RTTs tend to spread across three distinct ranges that are \SI{20}{\ms} and \SI{40}{\ms} apart, which causes step-wise lag distributions in Figs.~\ref{fig:lag_ukw:zoom} and \ref{fig:lag_ch:zoom}.  This suggests that Zoom may be employing \emph{regional load balancing} within the US when serving non-US sessions.  Whereas in Webex, RTTs to service endpoints consistently remain close to the trans-Atlantic RTTs~\cite{verizonstats}, indicating that non-US sessions are relayed via its infrastructure in US-east.  In case of Meet, its distributed service endpoints allow clients in Europe to enjoy stream lags that are comparable to US-based counterparts without any artificial detour.  The reason why its streaming lag is lower in Europe than in the US may be because the end-to-end latency among the clients (connected via service endpoints) in Europe may be smaller than that in the US. Average RTT within Europe is indeed smaller than that in the US~\cite{verizonstats}.

\subsection{User-Perceived Video Quality}
\label{sec:res:qoe:quality}
Next we shift our focus to user-perceived quality of videoconferencing.  A videoconferencing client typically captures a single person view against a stationary background, but it is also possible to have high-motion features in streamed content if a participant is joining a session from a mobile device, sharing a video playback with dynamic scenes, or showing a media-rich presentation, etc.
In general, however, little is known about how different videoconferencing systems measure up to one another in terms of user-perceived quality under different conditions.

\begin{figure*}[ht]
\centering
\begin{subfigure}{0.32\linewidth}
\centering
\includegraphics[width = 1.0\textwidth,trim = 5mm 1mm 5mm 9mm, clip=true]{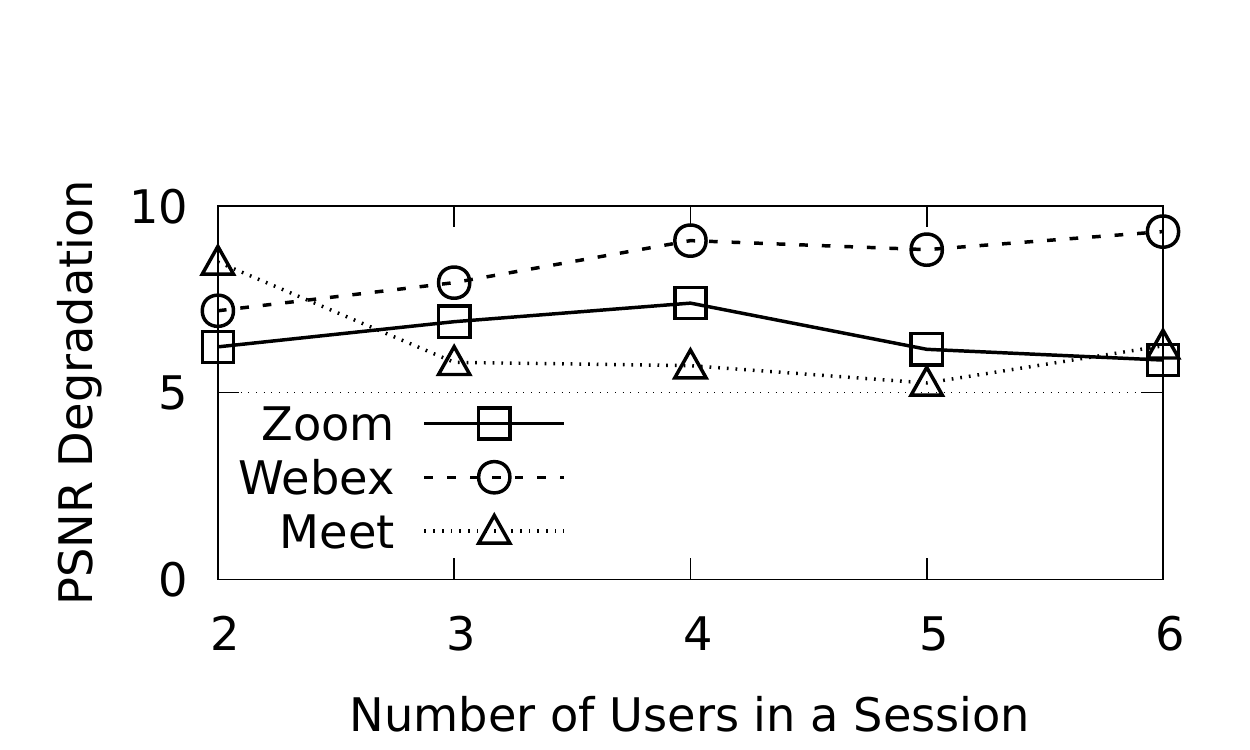}
\caption{PSNR}
\label{fig:psnr_diff_us}
\end{subfigure}
\begin{subfigure}{0.32\linewidth}
\centering
\includegraphics[width = 1.0\textwidth,trim = 5mm 1mm 5mm 9mm, clip=true]{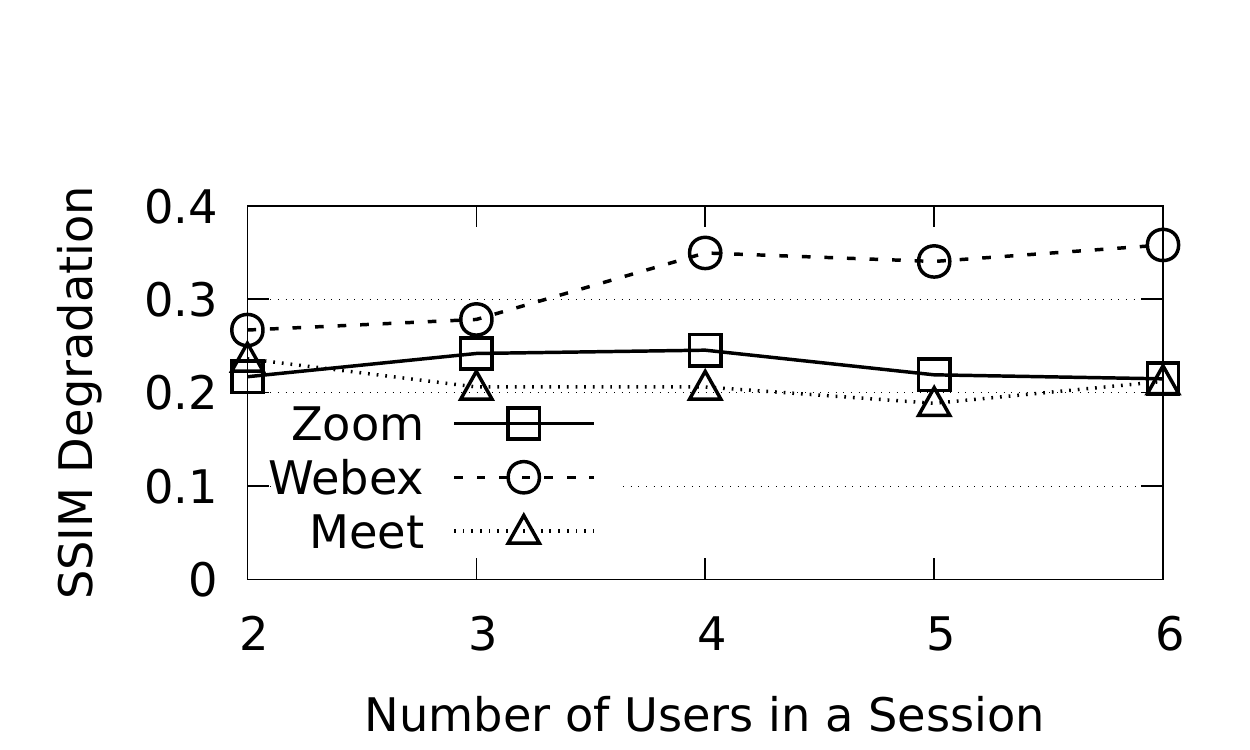}
\caption{SSIM}
\label{fig:ssim_diff_us}
\end{subfigure}
\begin{subfigure}{0.32\linewidth}
\centering
\includegraphics[width = 1.0\textwidth,trim = 5mm 1mm 5mm 9mm, clip=true]{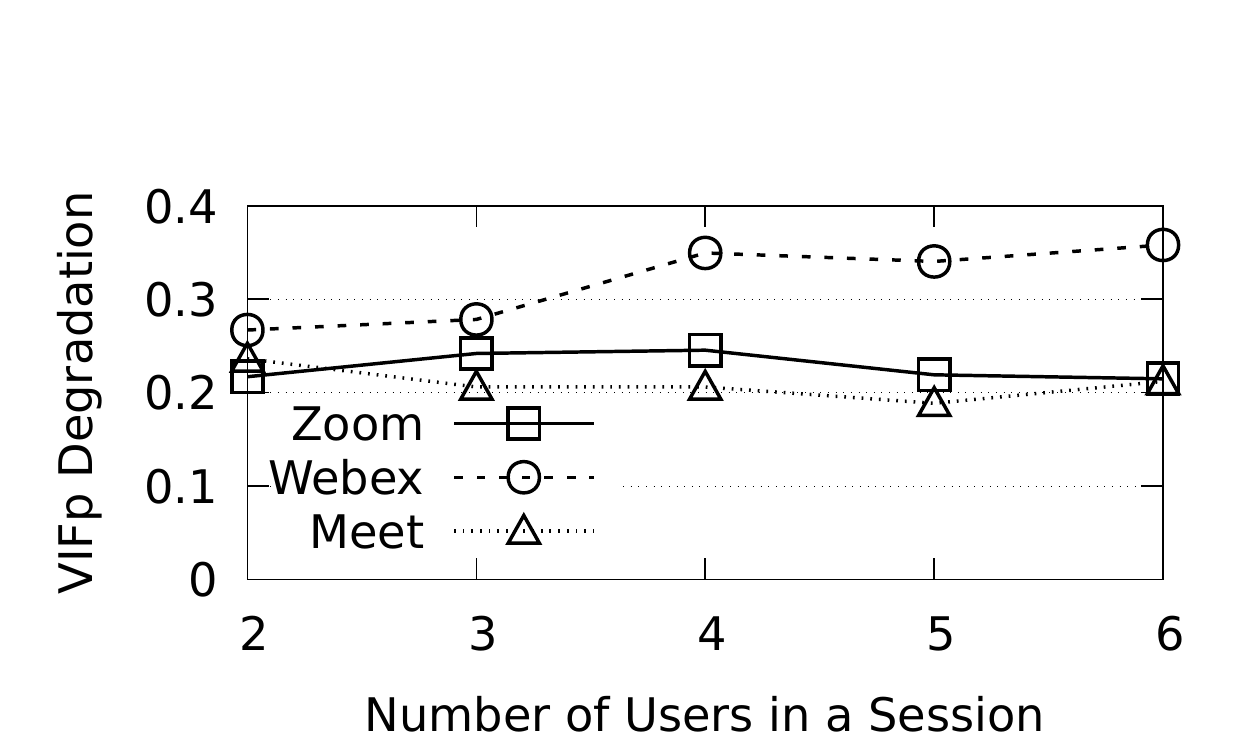}
\caption{VIFp}
\label{fig:vifp_diff_us}
\end{subfigure}
\vspace{-2ex}
\caption{Video QoE reduction when video feeds are changed from low-motion to high-motion (US).}
\label{fig:qoe_diff_us}
\vspace*{-1ex}
\end{figure*}

\begin{figure*}[ht]
\centering
\begin{subfigure}{0.45\linewidth}
\centering
\includegraphics[width = 1.0\linewidth,trim = 0mm 1mm 0mm 12mm, clip=true]{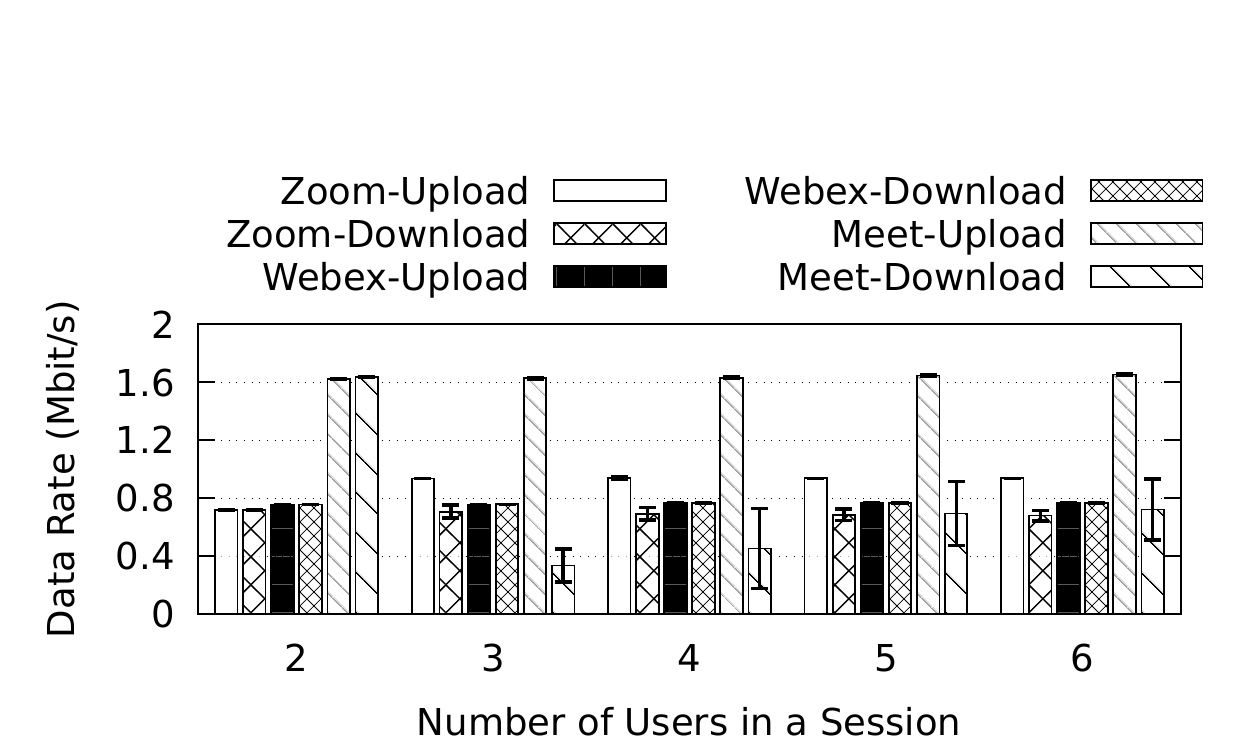}
\caption{US: low-motion}
\label{fig:datarate_us:lm}
\end{subfigure}
\begin{subfigure}{0.45\linewidth}
\centering
\includegraphics[width = 1.0\linewidth,trim = 0mm 1mm 0mm 12mm, clip=true]{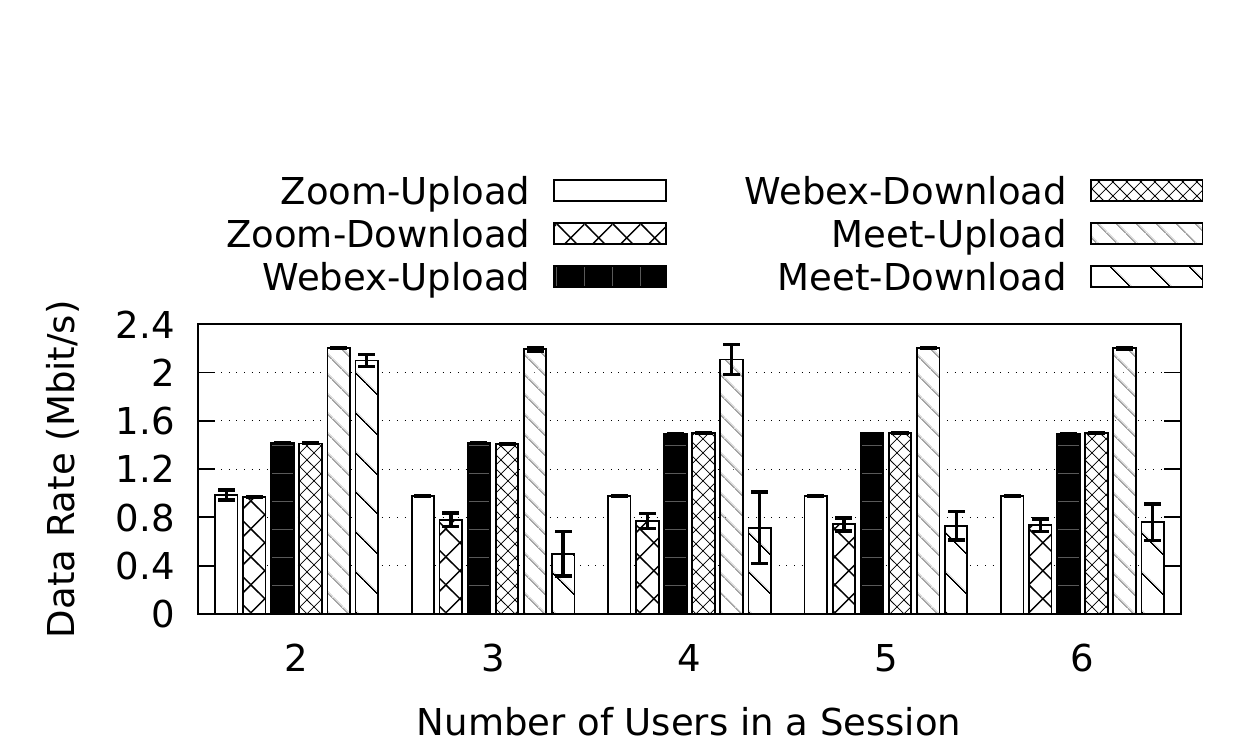}
\caption{US: high-motion}
\label{fig:datarate_us:hm}
\end{subfigure}
\vspace{-1ex}
\caption{Upload/download data rates (US). The data rate is computed from Layer-7 payload length in pcap traces. ``X-Upload'' shows the average upload rate of a meeting host (using Zoom, Webex, or Meet) who is broadcasting a video feed, while ``X-Download''  indicates the average download rate of the clients who are receiving the feed.  All uploads/downloads occur via cloud-based relay, except for Zoom with $N$=2 which uses peer-to-peer traffic.}
\label{fig:datarate}
\vspace*{-1ex}
\end{figure*}

\begin{figure*}[ht]
\centering
\begin{subfigure}{0.32\linewidth}
\centering
\includegraphics[width = 1.0\textwidth,trim = 5mm 1mm 5mm 5mm, clip=true]{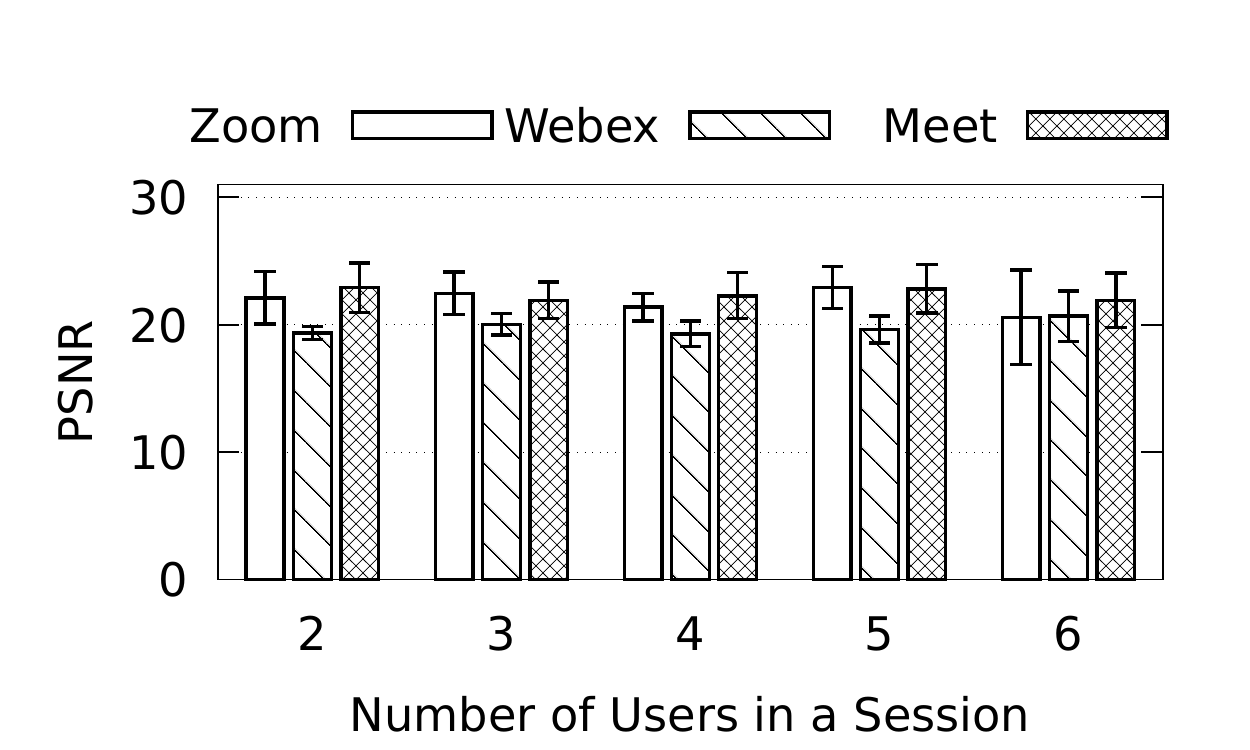}
\caption{High motion: PSNR}
\label{fig:scale_eu:hm:psnr}
\end{subfigure}
\begin{subfigure}{0.32\linewidth}
\centering
\includegraphics[width = 1.0\textwidth,trim = 5mm 1mm 5mm 5mm, clip=true]{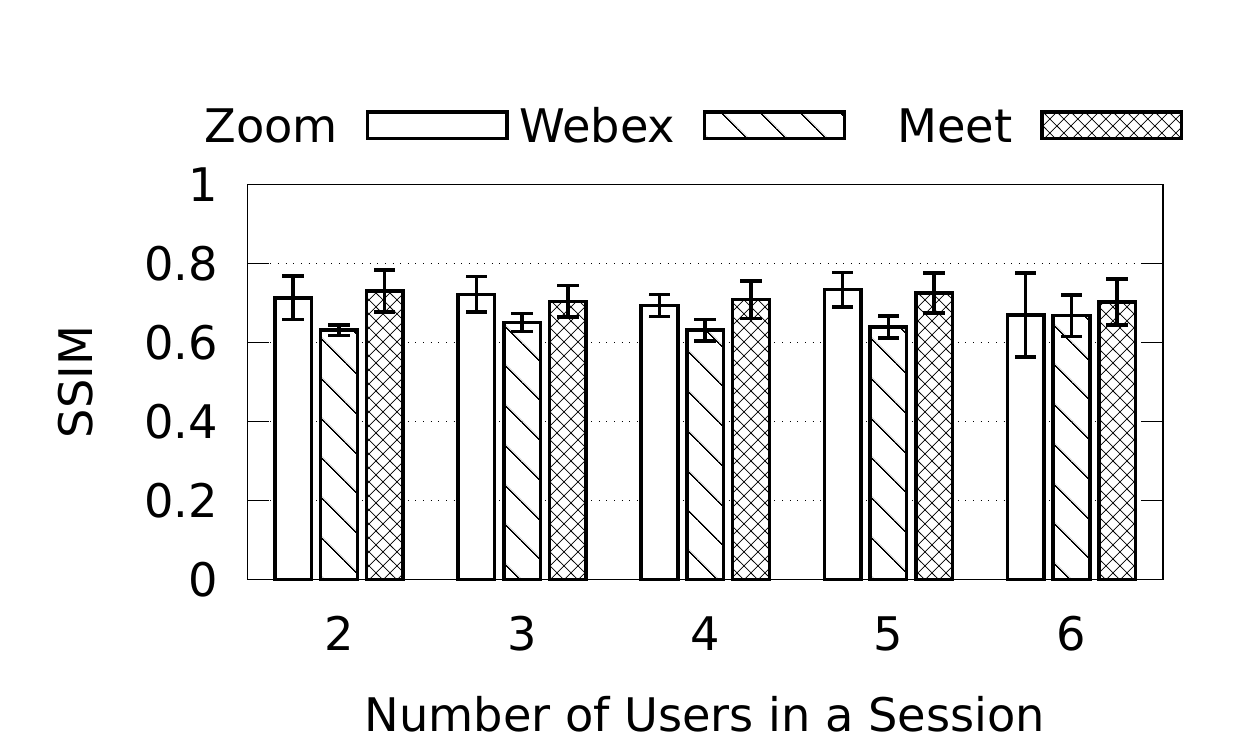}
\caption{High motion: SSIM}
\label{fig:scale_eu:hm:ssim}
\end{subfigure}
\begin{subfigure}{0.32\linewidth}
\centering
\includegraphics[width = 1.0\textwidth,trim = 5mm 1mm 5mm 5mm, clip=true]{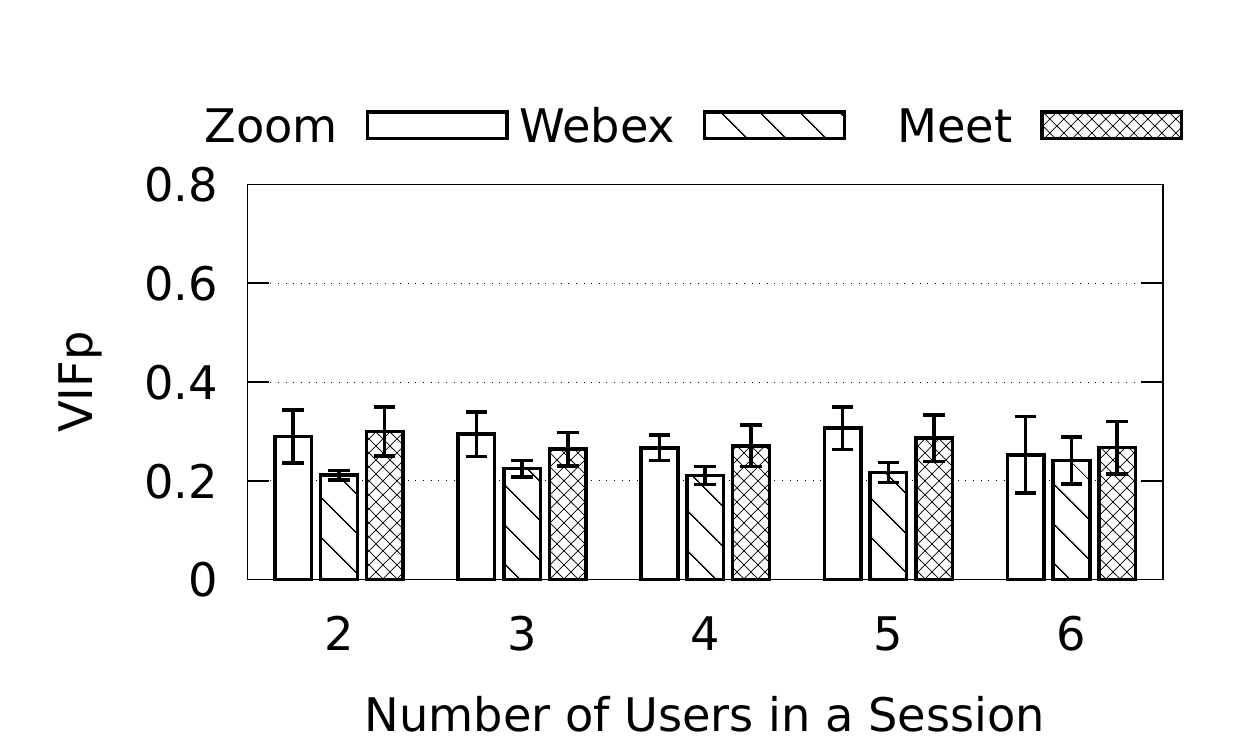}
\caption{High motion: VIFp}
\label{fig:scale_eu:hm:vifp}
\end{subfigure}
\vspace{-1ex}
\caption{Video QoE metrics comparison (Europe).}
\label{fig:scale_eu}
\vspace*{-2ex}
\end{figure*}

For this evaluation we prepare two distinct video feeds with 640$\times$480 resolution: (i) a low-motion feed capturing the upper body of a single person talking with occasional hand gestures in an indoor environment, and (ii) a high-motion tour guide feed with dynamically moving objects and scene changes.  On each videoconferencing system, we use a designated meeting host VM to create 10 five-minute long sessions, and inject the low-/high-motion videos into the sessions in an alternating fashion (hence two sets of five sessions).
In each session, we let $N$ clients join the session and render the received video feed in \emph{full screen} mode while their desktop screen is recorded locally.  For desktop recording, we use \texttt{PulseAudio} as audio backend, and set the video/audio codec to \texttt{H.264} (\SI{30}{fps}) and \texttt{AAC} (\SI{128}{Kbps}), respectively.  We repeat the whole experiment as we vary $N$ from one to five.\footnote{A typical number of users in a videoconferencing session is less than five~\cite{augusto21}.}

We compare the originally injected videos and recorded videos in terms of their quality with \texttt{VQMT}~\cite{vqmt}.  The \texttt{VQMT} tool computes a range of well-known objective QoE metrics, including PSNR (Peak Signal-to-Noise Ratio), SSIM (Structural Similarity Index Measure)~\cite{ssim} and VIFp (Pixel Visual Information Fidelity)~\cite{vifp}.  Each of these metrics produces \emph{frame-by-frame similarity} between injected/recorded videos. We take an average over all frames as a QoE value.  One issue that complicates accurate quality comparison is the fact that the video screen rendered by a client is partially blocked by client-specific UI widgets (\eg buttons, user thumbnails, etc.), even in full screen mode.  To avoid such partial occlusion inside the video viewing area, we prepare video feeds with enough padding (Fig.~\ref{fig:padding}).  When recorded videos are obtained, we perform the following post-processing on them before analysis.  We first crop out the surrounding padding and resize video frames to match the content layout and resolution of the injected videos. On top of that, we synchronize the start/end time of original/recorded videos with millisecond-level precision by trimming them in a way that per-frame SSIM similarity is maximized.

\subsubsection{US-based Videoconferencing.}
We use one cloud VM in US-east designated as a meeting host which broadcasts a stream, and up to five other VMs in US-west and US-east receiving the stream as passive participants.  Fig.~\ref{fig:scale_us} compares the quality of video streaming for these VMs in terms of three QoE metrics (PSNR, SSIM \& VIFp) as the number of users in a session (\emph{N}) increases. The height of bars indicates average QoE values across all sessions, with the errorbars being standard deviations. For easy comparison between low-motion and high-motion feeds, Fig.~\ref{fig:qoe_diff_us} shows the amount of QoE reduction with high-motion feeds (compared to low-motion feeds). Figs.~\ref{fig:datarate_us:lm} and~\ref{fig:datarate_us:hm} show the corresponding data rates for these sessions.

We make the following observations from the figures. Comparing Figs.~\ref{fig:scale_us:lm:psnr}--\ref{fig:scale_us:lm:vifp} against Figs.~\ref{fig:scale_us:hm:psnr}--\ref{fig:scale_us:hm:vifp}, one can find that, across all three platforms, low-motion sessions experience less quality degradation than high-motion sessions because their video feeds contain largely static background. The amount of decrease in QoE values between low-motion/high-motion sessions (Fig.~\ref{fig:qoe_diff_us}) is significant enough to downgrade mean opinion score (MOS) ratings by one level~\cite{moldovan16}.  On Webex, QoE degradation in high-motion scenario tends to become more severe with more users. Whereas no such consistent pattern is observed in Zoom and Meet.  The QoE results from low-motion sessions (Figs.~\ref{fig:scale_us:lm:psnr}--\ref{fig:scale_us:lm:vifp}) show that there is a non-negligible QoE drop between $N$=2 and $N$>2 on Meet. We find that, on Meet, the data rate for two-user sessions (\SIrange[range-phrase=--,range-units=single]{1.6}{2.0}{Mbps}) is significantly higher than other multi-user sessions (\SIrange[range-phrase=--,range-units=single]{0.4}{0.6}{Mbps}) (Fig.~\ref{fig:datarate}). Such higher traffic rate with $N$=2 helps with the QoE of low-motion sessions, but does not contribute much to the QoE of high-motion sessions. Among the three, Webex exhibits the most \emph{stable} QoE across different sessions. A similar behavior is observed in Figs.~\ref{fig:lag_useast}--\ref{fig:lag_ch}, where Webex shows the least variance in streaming lag as well.

\begin{figure*}[ht]
\centering
\begin{subfigure}{0.32\linewidth}
\centering
\includegraphics[width = 1.0\textwidth,trim = 5mm 0mm 5mm 18mm, clip=true]{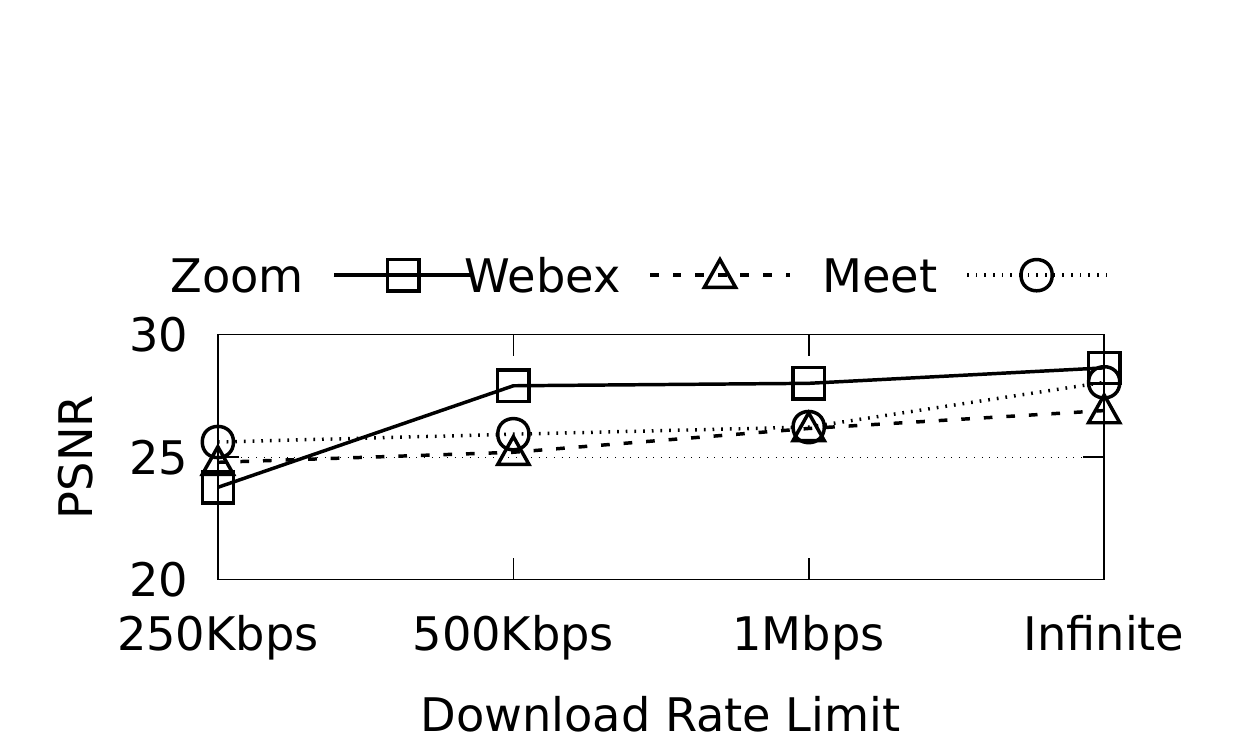}
\caption{Low motion: PSNR}
\label{fig:rl_psnr_low}
\end{subfigure}
\begin{subfigure}{0.32\linewidth}
\centering
\includegraphics[width = 1.0\textwidth,trim = 5mm 0mm 5mm 18mm, clip=true]{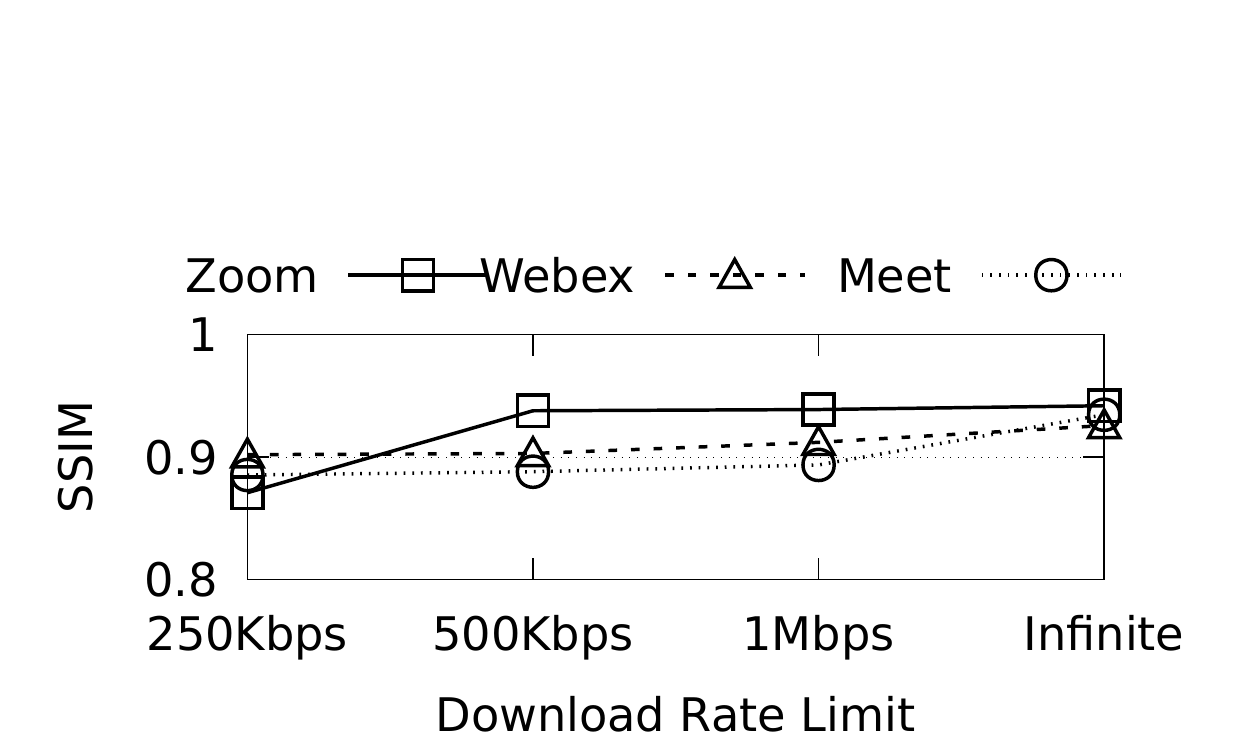}
\caption{Low motion: SSIM}
\label{fig:rl_ssim_low}
\end{subfigure}
\begin{subfigure}{0.32\linewidth}
\centering
\includegraphics[width = 1.0\textwidth,trim = 5mm 0mm 5mm 18mm, clip=true]{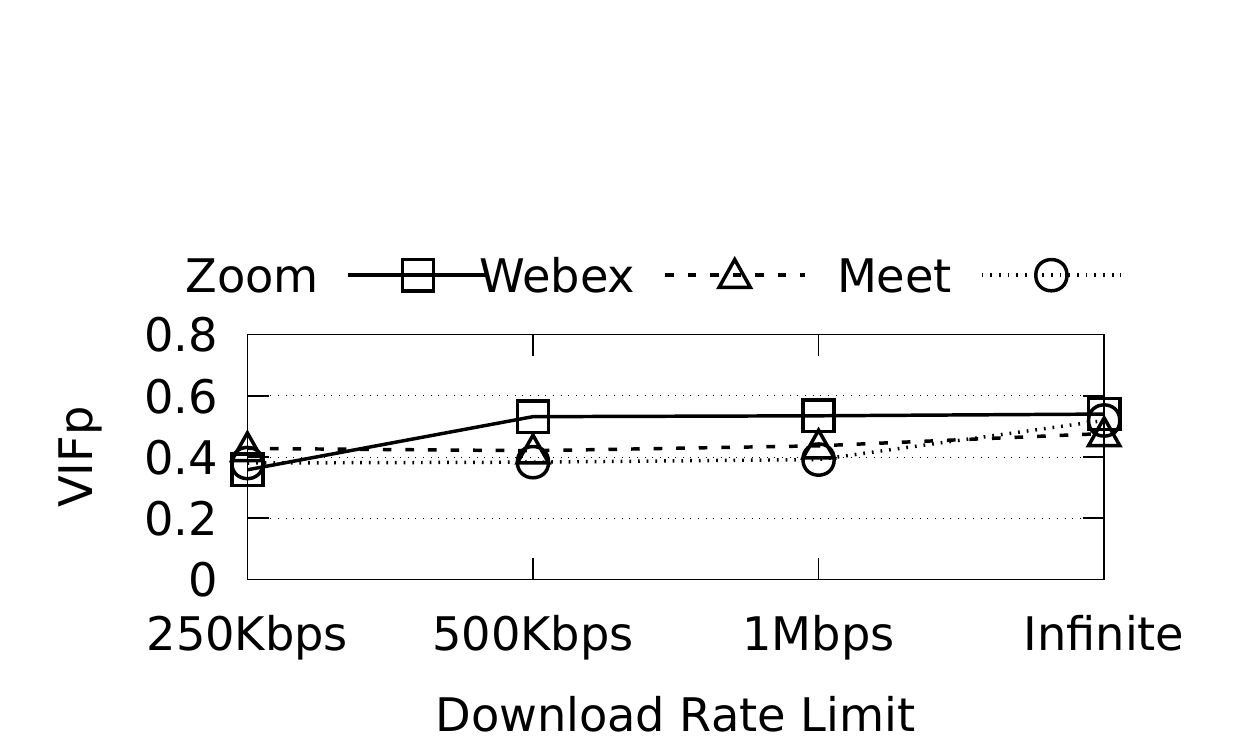}
\caption{Low motion: VIFp}
\label{fig:rl_vifp_low}
\end{subfigure}
\begin{subfigure}{0.32\linewidth}
\centering
\includegraphics[width = 1.0\textwidth,trim = 5mm 0mm 5mm 18mm, clip=true]{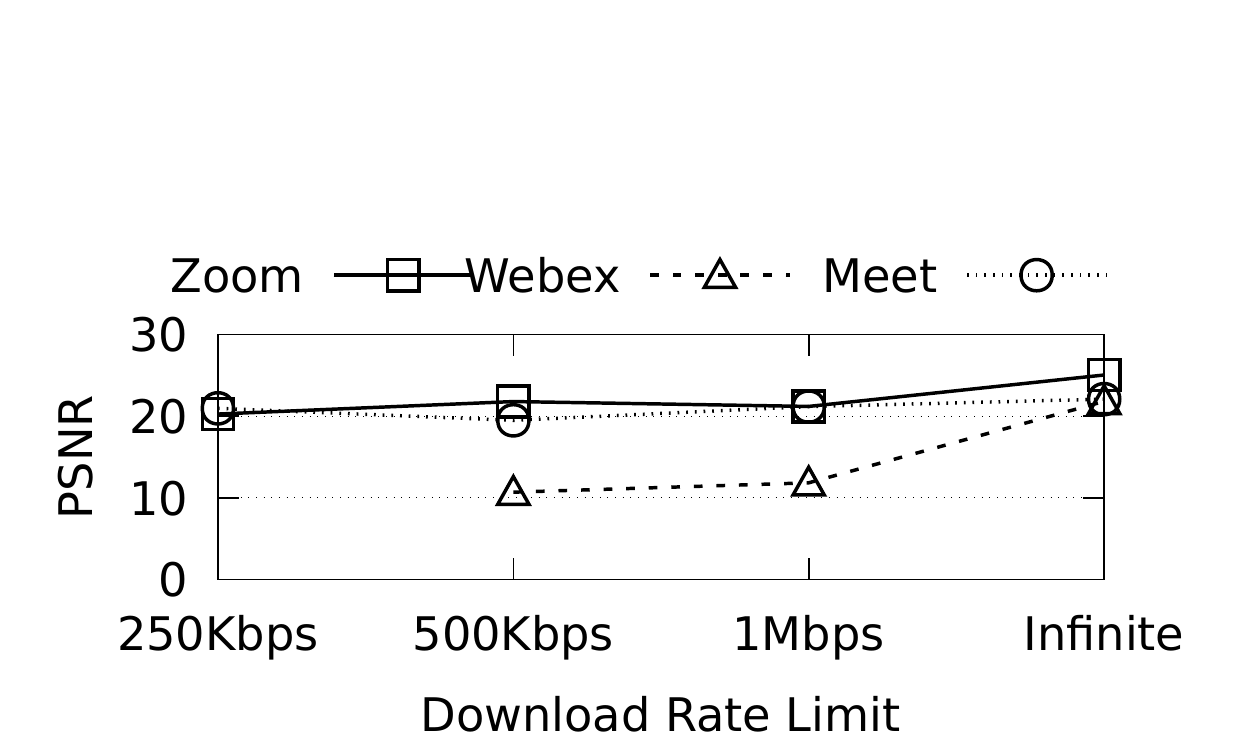}
\caption{High motion: PSNR}
\label{fig:rl_psnr_high}
\end{subfigure}
\begin{subfigure}{0.32\linewidth}
\centering
\includegraphics[width = 1.0\textwidth,trim = 5mm 0mm 5mm 18mm, clip=true]{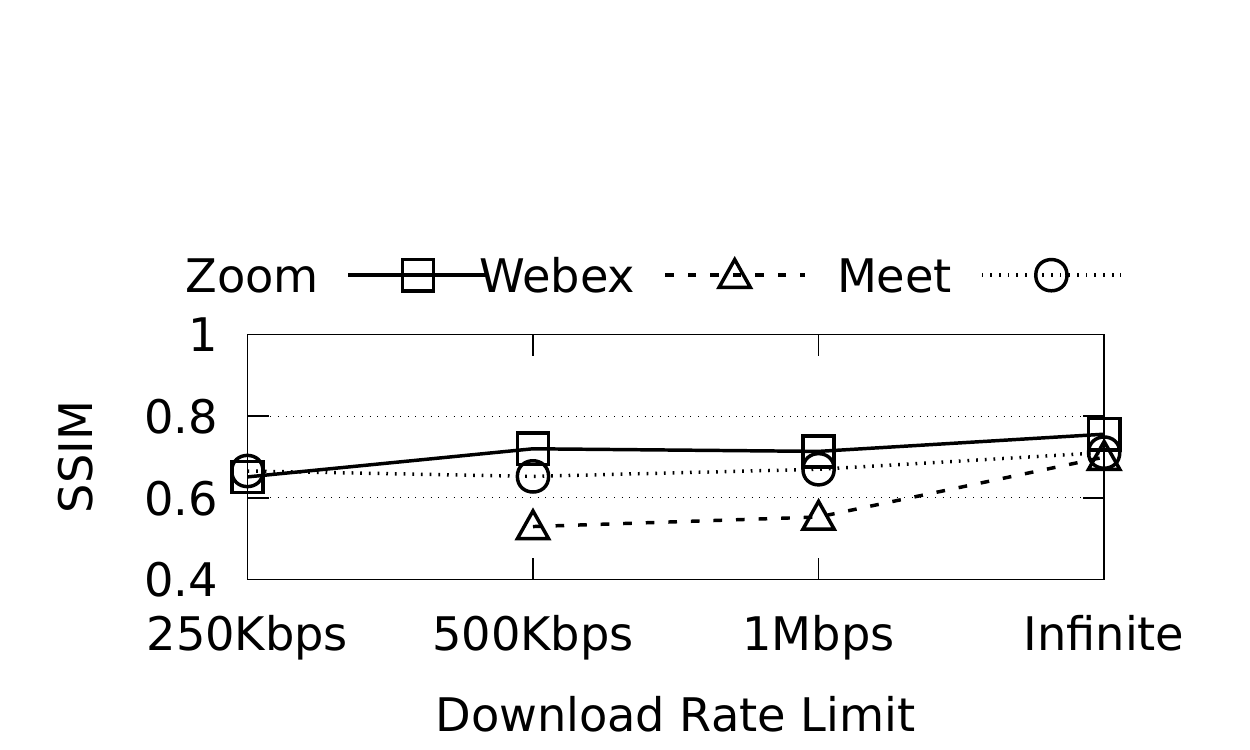}
\caption{High motion: SSIM}
\label{fig:rl_ssim_high}
\end{subfigure}
\begin{subfigure}{0.32\linewidth}
\centering
\includegraphics[width = 1.0\textwidth,trim = 5mm 0mm 5mm 18mm, clip=true]{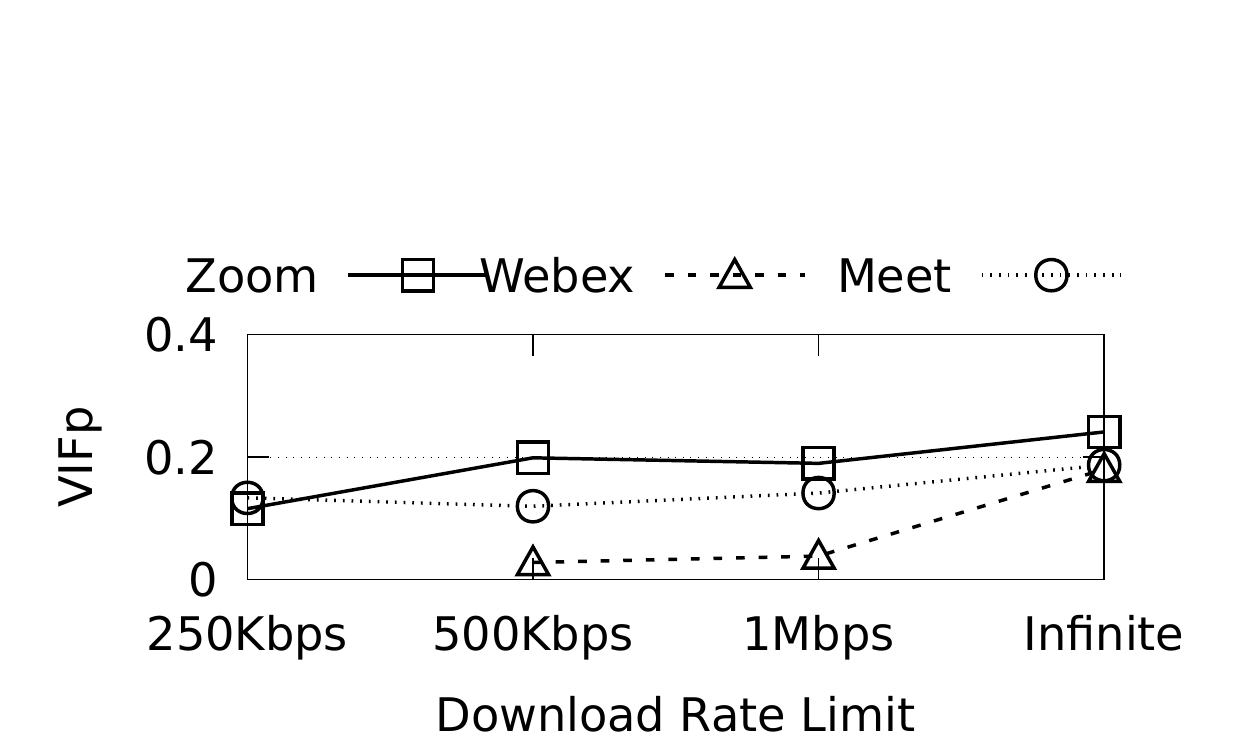}
\caption{High motion: VIFp}
\label{fig:rl_vifp_high}
\end{subfigure}
\vspace{-1ex}
\caption{Effect of bandwidth constraints on video quality.}
\label{fig:rl_qoe}
\vspace*{-1ex}
\end{figure*}

\begin{figure}[t]
\centering
\includegraphics[width = 0.9\linewidth,trim = 0mm 1mm 0mm 20mm, clip=true]{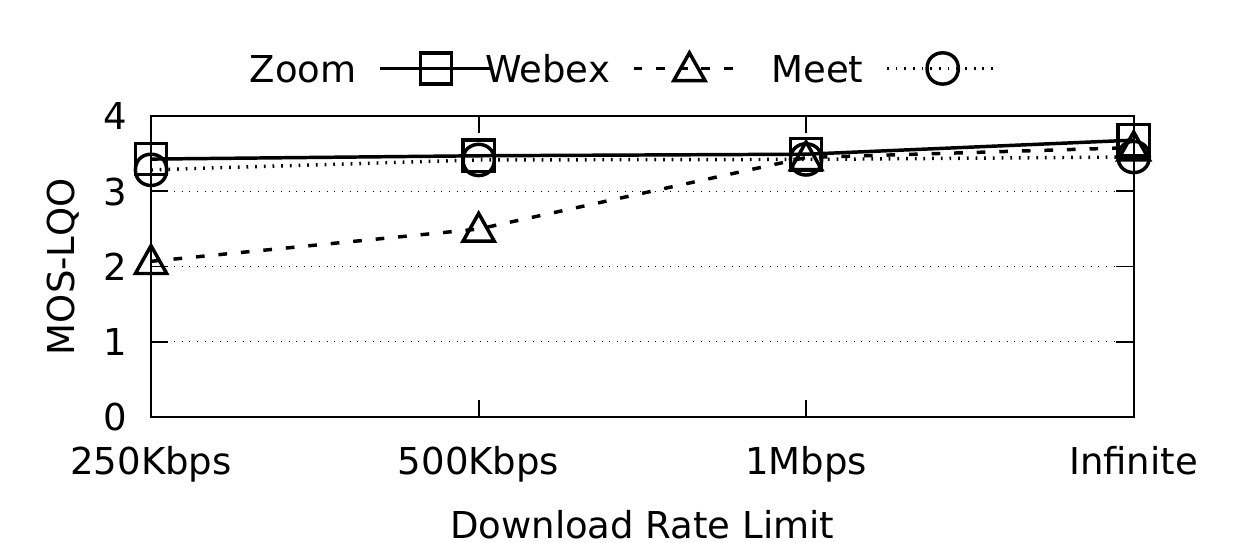}
\caption{Audio quality under bandwidth constraints. MOS-LQO is computed in speech mode for low-motion sessions which contain only human voice.}
\label{fig:audio_rl}
\vspace{-2ex}
\end{figure}

Traffic-wise (Fig.~\ref{fig:datarate}), we focus on two aspects: (1) data rate difference between low-motion and high-motion feeds, and (2) data rate variation across multiple sessions with the same feed.  All three systems send out a low-motion video feed in a lower rate than a high-motion counterpart as the former is more compressible.  The rate reduction in low-motion streams is the highest in Webex, where its low-motion sessions almost halve the required downstream bandwidth.  With a given video feed, Webex shows virtually no fluctuation in data rate across multiple sessions.  On the other hand, Meet reduces its data rate in low-motion sessions roughly by 20\% compared to high-motion sessions, but with much more \emph{dynamic} rate fluctuation across different sessions than Zoom/Webex.   Zoom exhibits the least difference (5--10\%) in data rate between low-motion and high-motion sessions. Its downstream data rate is slightly higher with peer-to-peer streaming ($\sim$\SI{1}{Mbps}; $N$=2) than with cloud-based relay ($\sim$\SI{0.7}{Mbps}; $N$>2).  When QoE metrics and data rates are considered together, Zoom appears to deliver the best QoE in the most bandwidth-efficient fashion, at least in the US.

\subsubsection{Non-US-based videoconferencing} We repeat the QoE analysis experiment using a set of VMs created in Europe, one VM in Switzerland designated as a meeting host, and up to five other VMs (in France, Germany, Ireland, UK) joining the session created by the host. Fig.~\ref{fig:scale_eu} shows QoE values on three systems with high-motion sessions.  The results from low-motion sessions are similar to those collected in the US, and thus are omitted.
When compared side-by-side, Meet maintains a slight edge in QoE metrics among three systems, potentially due to its European presence.  In case of Zoom, although its average QoE appears to be similar to that of Meet, its QoE variation across different sessions is higher than Meet for high $N$. This observation is aligned with our earlier finding in Fig.~\ref{fig:lag_ukw:zoom}, where we show that its regional load balancing causes more variable streaming lag in Europe.

\subsection{Streaming under Bandwidth Constraints}
\label{sec:res:qoe:bw}
The experiments presented so far are conducted in an \emph{unlimited} bandwidth environment. The cloud VMs used have a bidirectional bandwidth of multi-Gbps~\cite{azure_fsv2}, which far exceeds the measured data rates of \SIrange[range-phrase=--,range-units=single]{1}{2}{Mbps} (Fig.~\ref{fig:datarate}). In the next set of experiments, we apply artificial bandwidth caps on our cloud VM and examine its effect on QoE.  We use Linux \texttt{tc}/\texttt{ifb} modules to enable traffic shaping on incoming traffic.  Here we present QoE metric analysis not just for video but also for audio. We extract video and audio data separately from recorded videoconferencing sessions.  For audio QoE analysis, we perform the following processing on extracted audio.  First, we normalize audio volume in the recorded audio (with EBU R128 loudness normalization), and then synchronize the beginning/ending of the audio in reference to the originally injected audio.  We use the \texttt{audio-offset-finder} tool for this.  Finally, we use the \texttt{ViSQOL} tool~\cite{visqol} with the original/recorded audio data to compute the MOS-LQO (Mean Opinion Score - Listening Quality Objective) score, which ranges from 1 (worst) to 5 (best).

Figs.~\ref{fig:rl_qoe} and \ref{fig:audio_rl} show how video/audio QoE metrics change under various rate-limiting conditions.  Each dot in the figures represents the average QoE values of five 5-minute long sessions.

\vspace{0.1in}
\noindent\textbf{Video QoE.} Overall, Zoom tends to maintain the best QoE with decreasing bandwidth limits, although there is sudden drop in QoE with a bandwidth cap of \SI{250}{Kbps}.  Meet maintains more graceful QoE degradation across all scenarios.  Webex suffers from the most significant QoE drops with smaller bandwidth caps.  With bandwidth $\leq$ \SI{1}{Mbps}, video frequently stalls and even completely disappears and reappears on Webex.

\vspace{0.1in}
\noindent\textbf{Audio QoE.} Compared to non-negligible QoE drops in video, audio QoE levels remain virtually constant on Zoom and Meet, probably due to the low data rate of audio (\SI{90}{Kbps} for Zoom and \SI{40}{Kbps} for Meet).\footnote{We measure their audio rates separately using audio-only streams.} However, voice quality on Webex, even with its low rate (\SI{45}{Kbps}), is relatively sensitive to bandwidth limits, starting to deteriorate noticeably (e.g., manifested as distorted/paused sound) with a limit of \SI{500}{Kbps} or less.


\begin{figure*}[ht]
\centering
    \begin{subfigure}{0.32\linewidth}
    \centering
    \includegraphics[width = 1.0\textwidth,trim = 5mm 5mm 5mm 0mm, clip=true]{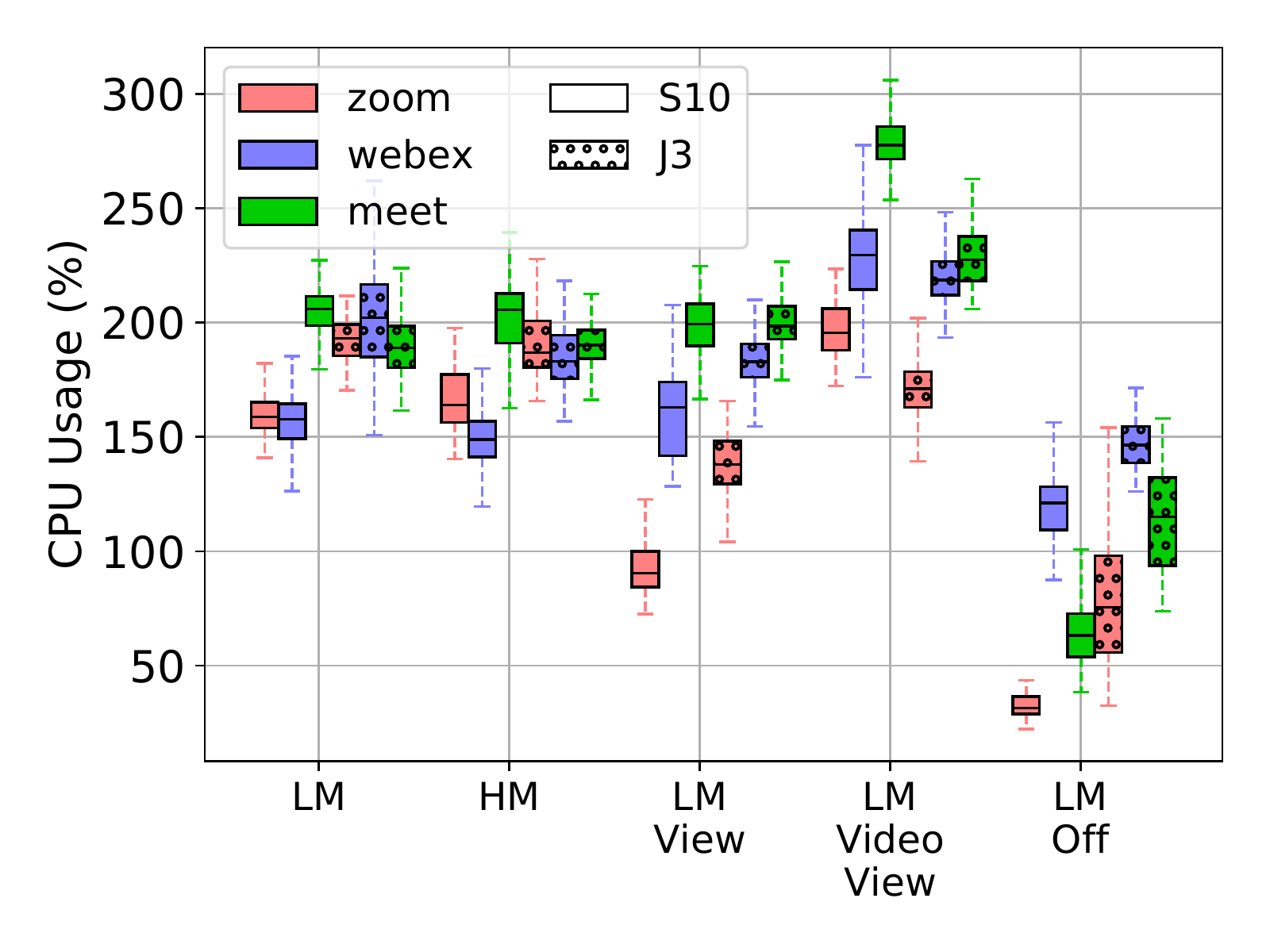}
    \caption{CPU usage}
    \label{fig:res:android:cpu}
    \end{subfigure}
    \begin{subfigure}{0.32\linewidth}
    \centering
    \includegraphics[width = 1.0\textwidth,trim = 5mm 5mm 5mm 0mm, clip=true]{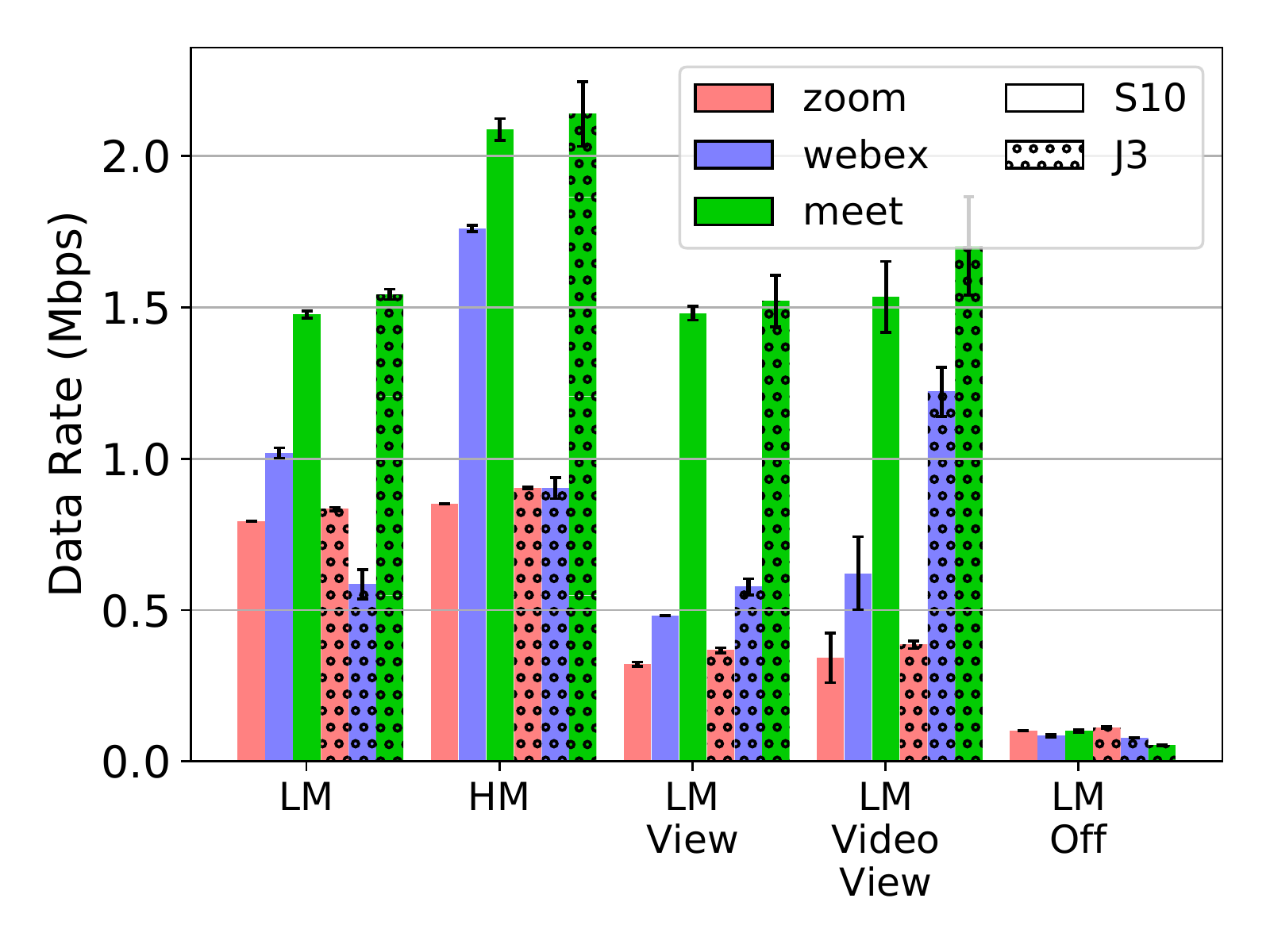}
    \caption{Data rate}
    \label{fig:res:android:bdw}
    \end{subfigure}
    \begin{subfigure}{0.32\linewidth}
    \centering
    \includegraphics[width = 1.0\textwidth,trim = 5mm 5mm 5mm 0mm, clip=true]{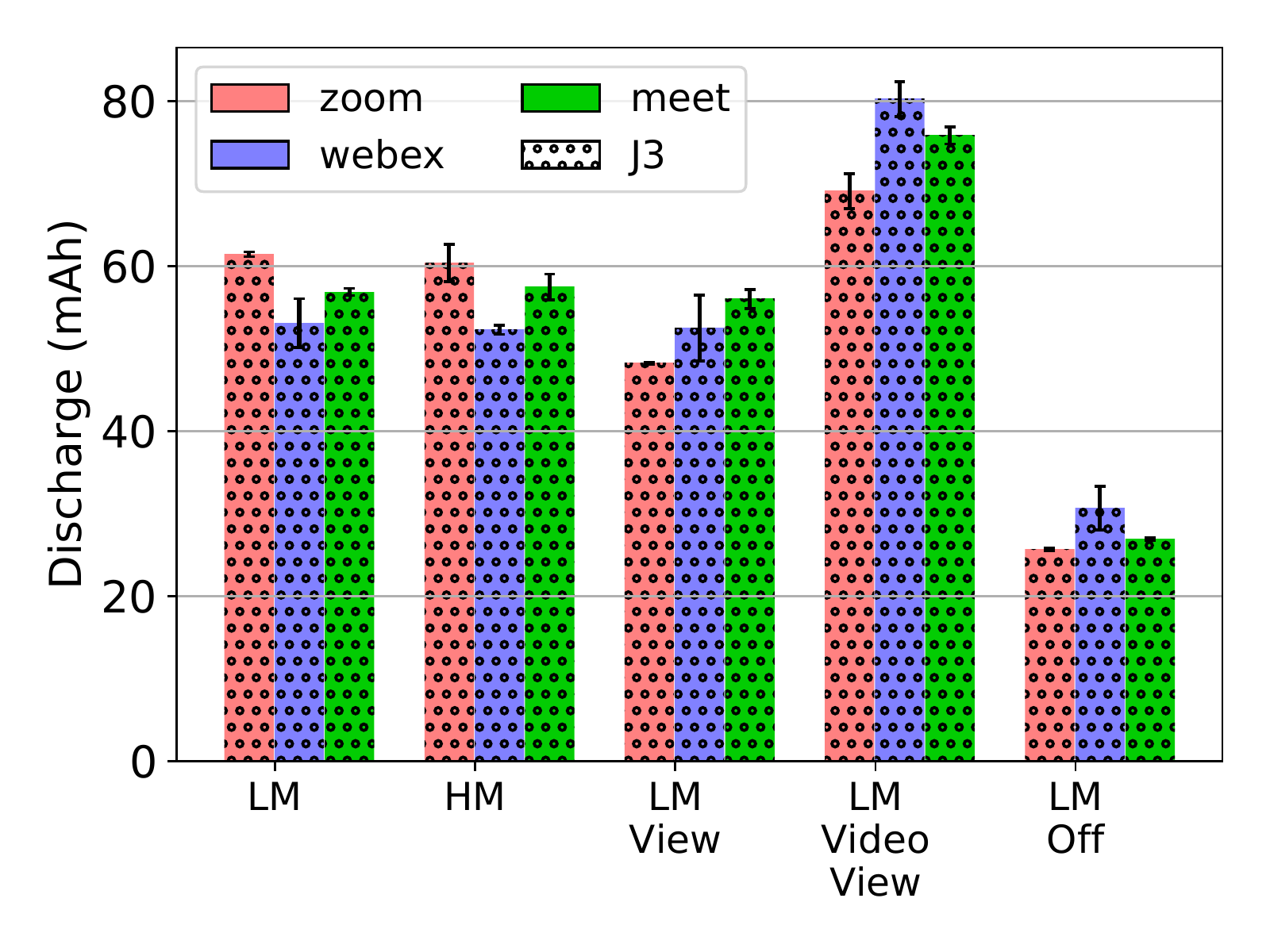}
    \caption{Battery discharge (J3 only)}
    \label{fig:res:android:batt}
    \end{subfigure}
\caption{Resource consumption evaluation. Samsung S10 and J3 (Android).}
\label{fig:res:android}
\end{figure*}
\section{Resource Consumption}
\label{sec:res:usage}
After having investigated user-perceived QoE offered by three major videoconferencing systems, we shift our attention to their client-side \textit{resource consumption}, such as CPU, bandwidth and battery usages. For this analysis, we resort to two Android devices: S10 (high-end) and J3 (low-end) as described in Table~\ref{tab:testbed}. In addition to these devices, each experiment includes one cloud VM designated as a meeting host. Since the Android devices are located in a residential access network of the east-coast of US, we run the meeting host in a US-east VM. We set the videoconference duration to five minutes, and repeat each experiment five times. The meeting host streams the two previously introduced video feeds: low-motion (\texttt{LM}) and high-motion (\texttt{HM}). 


At the Android devices, we consider several device/UI settings which can affect videoconferencing sessions.  Unless otherwise noted by the label ``\texttt{Video}'', each device's camera is turned off to minimize noise, and the incoming video feed is displayed in full screen. Given the low-motion video, we: 1) change the videoconferencing client's view into a gallery-view (\texttt{LM-View}), which assigns to each videoconference participant an equal portion of the screen,\footnote{Meet has no support for this feature. We approximate it by ``zooming out'', \ie revealing UI controls like mute or hang up.}  2) turn on the cameras and the gallery-view (\texttt{LM-Video-View}), and 3) turn off both camera and screen (\texttt{LM-Off}), simulating a driving scenario. 

\vspace{0.1in}
\noindent\textbf{CPU usage.} Fig.~\ref{fig:res:android:cpu} shows boxplots of CPU usage sampled every three seconds for the experiment duration across all devices and scenarios. 
Each boxplot accounts for CPU samples collected across five repetitions of an experiment. We report CPU usage in absolute numbers, \eg 200\% implies full utilization of two cores. If we focus on the \texttt{LM} and \texttt{HM} scenarios for S10 (high-end), the figure shows that Zoom and Webex have comparable CPU usage (median of 150--175\%), while Meet adds an extra 50\%. When we focus on J3 (low-end device), CPU usage among the three clients is instead comparable (median around 200\%). This indicates a dynamic behavior of the Meet client which only grabs more resources if available.

Zoom is the only client which benefits from the gallery view (both when the device's camera is on or off), reducing its CPU usage by 50\% on both devices. Meet sees no benefit from this setting, which is expected given that it has no direct support for it, \ie the meeting host's video still occupies about 80\% of the screen.  Surprisingly, Webex does not benefit from its gallery view, even causing a slight CPU increase on S10. Irrespective of the videoconferencing client, activating the device's camera (\texttt{LM-Video-View}) adds an extra 100\% and 50\% of CPU usage on S10 and J3, respectively. The higher CPU usage on S10 is due to a better camera with twice as many megapixels (10M), HDR support, etc. 

Finally, CPU usage is minimized when the device screen is off. However, while Zoom and Meet's CPU usage is reduced to 25--50\% (S10), Webex still requires about 125\%. This result, coupled with the lack of benefit of Webex's gallery setting, indicates some room for Webex to improve their Android client with more careful detection of user settings, as achieved by its competitors.  


\vspace{0.1in}
\noindent\textbf{Data rate.} We now focus on the Layer-7 download data rate, computed directly from pcap traces. Fig.~\ref{fig:res:android:bdw} shows the average download data rate per client, device, and setting, with errorbars reporting the standard deviation. If we focus on the high-end device (S10) and the \texttt{LM}/\texttt{HM} scenarios, the figure shows a trend similar to Fig.~\ref{fig:res:android:cpu}: Zoom uses the lowest data rate while Meet the highest, and the HM video causes a significant data rate increase, with the exception of Zoom. When focusing on the low-end device (J3), we notice that only Webex shows a ``truly'' adaptive behavior, \ie lower data rate for both LM and the low-end device. Zoom instead sticks to a somehow default data rate (\SI{750}{Kbps}), while Meet only takes into account the quality of the video, and not the target device. 
As previously observed for CPU usage, Meet's data rate is not impacted by the client-side gallery view, while it drops Zoom's data rate by 50\%, both with a device's video on and off. Webex's gallery view is instead less data efficient, particularly when a device's video is on. In this case, J3 reports a significant increase in the data rate (more than doubled compared to the \texttt{LM} scenario) due to the video sent by S10. In comparison, S10 reports a much lower video rate (\SI{700}{Kbps} vs.~\SI{1.2}{Mbps}) due to the J3's lower quality camera, as well as lack of light due to its location in the lab. Finally, the \texttt{LM-Off} scenarios confirm that no video is streamed when the screen is off, and just \SIrange[range-phrase=--,range-units=single]{100}{200}{Kbps} are needed for audio. 

\vspace{0.1in}
\noindent\textbf{Battery usage.} Next, we report on the battery consumption associated with videoconferencing. In this case, we only focus on J3, whose (removable) battery is connected to a Monsoon power meter~\cite{monsoon}. Fig.~\ref{fig:res:android:batt}  confirms that videoconferencing is an expensive task on mobile, draining up to 40\% of its 2600mAh battery during an one-hour conference with camera on. Overall, the figure shows no dramatic difference among the three clients, whose battery usage is within 10\% of each other. Zoom is the most energy efficient client with gallery view (\texttt{LM-View}), which provides a 20\% reduction compared with \texttt{LM}. Gallery view does not provide benefits to both Webex and Meet, similar to what we reported for both CPU usage and data rate. Finally, turning off the screen and relying on audio only saves up to 50\% of a user's battery.


\vspace{0.1in}
\noindent\textbf{Videoconference size.} Finally, we investigate the impact of the number of conference participants on client-side resource utilization. To do so, in addition to the meeting host, we introduce up to eight cloud-VMs as participants. To further stress the devices under test, we configure the host as well as the extra eight cloud VMs to stream a high-motion video simultaneously. We consider two UI settings in the Android clients: \textit{full screen} and \textit{gallery}. 

\begin{table}[t]
\scriptsize
\centering
\setlength\tabcolsep{3pt}
\caption{Data rate and CPU usage with various videconference sizes ($N$). Each cell reports statistics for S10/J3.}
\vspace{-2ex}
\label{tab:comp}
\begin{tabular}{C{0.5cm} | C{0.9cm} | C{1.7cm} C{1.1cm} | C{1.7cm} C{1.1cm}}\hline
\textbf{N} & \textbf{Client}  & \multicolumn{2}{c|}{\textbf{Full screen}} & \multicolumn{2}{c}{\textbf{Gallery}} \\\hline
  &    &  \textbf{Data rate (Mbps)} & \textbf{CPU (\%)}  & \textbf{Data rate (Mbps)} & \textbf{CPU (\%)} \\\hline
\multirow{3}{*}{3} & \textbf{Zoom}      & 0.85/0.9    & 164/186   & 0.33/0.37 & 102/148 \\
                     &  \textbf{Webex}     & 1.76/0.9    & 148/183   & 0.57/0.59 & 149/186 \\
                    & \textbf{Meet}      & 2.08/2.13   & 205/190   & 2.08/2.11 & 209/200 \\\hline
\multirow{3}{*}{6} & \textbf{Zoom}      & 0.92/0.94   & 189/212   & 0.71/0.73  & 101/152  \\
                     & \textbf{Webex}     & 1.75/0.9    & 140/195   & 0.43/0.47  & 155/184  \\
                     & \textbf{Meet}      & 2.25/2.33   & 257/211   & 2.15/2.24  & 235/219  \\\hline
\multirow{3}{*}{11} & \textbf{Zoom}      & 0.91/0.96   & 191/211   & 0.73/0.75  & 100/150  \\
                      & \textbf{Webex}     & 1.76/0.89   & 141/194   & 0.48/0.43  & 154/182  \\
                      & \textbf{Meet}      & 2.24/2.36   & 258/210   & 2.16/2.26  & 236/220  \\\hline
\end{tabular}
\end{table}


Table~\ref{tab:comp} summarizes the results; we report on average data rate and median CPU utilization per device (S10/J3, in each cell) and scenario. Note that $N=3$ is the scenario we have evaluated so far, \ie one cloud VM as a meeting host plus two Android devices. In full screen mode, Meet incurs a significant increase in both data rate (10\%) and CPU usage (50\%). This is because, even in full screen, Meet still shows a small preview of the video of the other two participants (plus the device's video, if on). Conversely, Zoom and Webex appear visually equivalent in full screen regardless of $N$. Still, for Zoom we observe a small increases in its data rate (5\%) and CPU (12\%) suggesting that some additional video streams are being buffered in the background with the goal to minimize latency in case a user decides to change the  view into \textit{gallery}. Although we could not capture such latency, we visually confirmed that both Zoom and Meet allow us to rapidly switch between participants, while Webex incurs some buffering time in the order of seconds. 

If we focus on the gallery view, we see that the extra participants cause a twofold data rate increase for Zoom, but no additional CPU usage. A similar trend is observed for Webex CPU-wise, but in this case we also observe a counter-intuitive data rate reduction (from \SI{600}{Kbps} down to \SI{450}{Kbps}).  Upon visual inspection, we find that this reduction is associated with a significant quality degradation in the video stream. Further increasing the participants to 11 does not lead to additional resource consumption. This happens  because the gallery view of Zoom and Webex -- as well as the only Meet's setting --   show videos for up to four concurrent participants.

\section{Limitations}
\label{sec:disc}
We conclude the paper by discussing several limitations of our study and future works we plan to explore.

\vspace{0.05in}
\noindent\textbf{Effect of last mile.} Our cloud-based vantage points may not represent the realistic last-mile network environments (e.g., broadband, wireless) of typical videoconferencing users from two perspectives.  First, the upstream connectivity of cloud-emulated clients (with multi-Gbps available bandwidth) is too idealistic.  While our experiments with bandwidth emulation (Section~\ref{sec:res:qoe:bw}) and a mobile testbed (Section~\ref{sec:res:usage}) are intended to address this limitation, a more realistic QoE analysis would consider dynamic bandwidth variation and jitter as well.  Another caveat is that all our emulated clients are created inside a \emph{single} provider network (i.e., Azure network). Thus any particular connectivity/peering of the Azure network might influence our cloud experiments.  Ideally, the testbed should be deployed across multiple cloud providers to mitigate any artifact of a single provider, or even encompass distributed edge-based platforms provisioned across heterogeneous access networks (e.g., residential~\cite{mcc14,batterylab2019}, campus~\cite{cloudlab} and enterprise networks~\cite{azureedge,awssnow}).  In fact, moving the evaluation platform to the edge would allow us to extend the list of target videoconferencing systems to study.\footnote{The use of Azure cloud prevents us from including Microsoft Team in our evaluation list.}

\vspace{0.05in}
\noindent\textbf{Free-tier vs.~paid subscription.} The validity of our findings is limited to the free-tier services.  Given the prevalent multi-CDN strategies~\cite{bitag} and differentiated product offerings, user's QoE can change under different subscription plans or any special license arrangements.  For example, Zoom is known to allow users with a university license to connect to AWS in Europe~\cite{sandervideo}, which we do not observe with its free-tier/paid subscription plans.  In case of Webex, we confirm that, with a paid subscription, its clients in US-west and Europe can stream from geographically close-by Webex servers (with RTTs $<$ \SI{20}{ms}).  Our methodology allows us to easily extend the scope of our study beyond the free-tier services.

\vspace{0.05in}
\noindent\textbf{Videoconferencing scalability.} Our QoE analysis targets small-size videoconferencing sessions (with up to 11 participants). An interesting question is how well user's QoE on each system scales as a videoconferencing session is joined by a moderate/large number of participants.  One possible approach would use a mix of crowd-sourced human users (who would generate varied-size videoconferencing sessions) and cloud VMs (which would be under our control for detailed QoE analysis of such sessions).

\vspace{0.05in}
\noindent\textbf{Black-box testing.} Our measurement study is a case of \textit{black-box testing}, where we do not have any knowledge on inner workings of individual systems we study. As such we are severely limited in our ability to \textit{explain} some of the observations we are making.  That said, we still argue that our study presents a valuable contribution to the community.  For one, our platform-agnostic methodology is general enough for any arbitrary videoconferencing systems.  Also, the proposed evaluation scenarios can be a useful input to videoconferencing operators for enhancing their infrastructures and clients.

\vspace{0.05in}
\noindent\textbf{Videoconferencing client.} Our emulated client runs on Linux only (via in-kernel virtual devices).  We consider that the modern Linux environment is representative enough for videoconferencing due to the good Linux support~\cite{zoomclient} or the use of cross-platform web clients by the existing systems.  For completeness, one can extend the client emulation beyond Linux, at least in a desktop environment, using similar device emulation and workflow automation tools on Windows~\cite{obscam-win,powerautomate} and MacOS~\cite{obscam-macos,automator}.  The QoE analysis could then be extended to cover different mobile and desktop environments in a more comprehensive fashion.


\bibliographystyle{ACM-Reference-Format}
\bibliography{biblio}


\end{document}